\documentclass[journal, a4paper]{IEEEtran}

\usepackage{amsmath}
\usepackage{amssymb}
\usepackage[tight]{subfigure}
\usepackage[small]{caption} 
\usepackage{amsfonts}
\usepackage{graphics}
\usepackage{epsfig}
\usepackage{psfrag}
\usepackage{pstricks}
\usepackage{array}
\usepackage{shortvrb}
\usepackage{epsf}
\usepackage{graphicx}
\usepackage{rotating}
\usepackage{float}
\usepackage{color}
\usepackage{cite}
\usepackage{soul}

\usepackage{flushend}

\usepackage[left=1.5cm,top=2.0cm,right=1.5cm,bottom=1.75cm]{geometry} 

\graphicspath{{./eps/}}
\DeclareGraphicsExtensions{.eps}

\def \R {{\rm I\kern -2.2pt R\hskip 1pt}}

\newcommand{\PreserveBackslash}[1]{\let\temp=\\#1\let\\=\temp}
\newcommand{\bfg}[1]{\boldsymbol{#1}}

\begin{document}
\title{Generalized Model of VSC-based Energy Storage Systems for
  Transient Stability Analysis}
\author{{\'A}lvaro Ortega,\thanks{{\'A}.~Ortega and F.~Milano are with
    the School of Electrical, Electronic and Communications
    Engineering of the University College Dublin, Dublin, Ireland. \newline
    (e-mails: alvaro.ortega-manjavacas@ucdconnect.ie;
    federico.milano@ucd.ie).} {\em Student Member,
    IEEE}, Federico Milano, {\em Senior Member, IEEE} 
\vspace{-5mm} }

%
%
\maketitle

\begin{abstract}
  This paper presents a generalized energy storage system model for
  voltage and angle stability analysis. The proposed solution
    allows modeling most common energy storage technologies through a
    given set of linear differential algebraic equations (DAEs). 
    In particular, the paper considers, but
  is not limited to, compressed air, superconducting magnetic,
  electrochemical capacitor and battery energy storage devices.  While able
  to cope with a variety of different technologies, the proposed
  generalized model proves to be accurate for angle and voltage
    stability analysis, as it includes a balanced,
  fundamental-frequency model of the voltage source converter (VSC)
  and the dynamics of the dc
  link.  Regulators with inclusion of hard limits are also taken into
  account.  The transient behavior of the generalized model is
  compared with detailed fundamental-frequency balanced
  models as well as commonly-used simplified models of
  energy storage devices.  A comprehensive case study based on the
  WSCC 9-bus test system is presented and discussed.
\end{abstract}


\vspace{-2.5mm}

\section{Introduction}
\label{sec:intro}



The capability of Energy Storage Systems (ESSs) to improve the
transient behavior of power systems and to increase the
competitiveness of non-dispatchable power production technologies,
e.g., renewable energies, has led, in recent years, to a huge
investment in research, prototyping and installation of storage
devices as well as the development of a huge variety of energy storage
technologies~\cite{ibrahim:08, beaudin:10, IEC:2011}.  Among most
relevant technologies that currently appear most promising there are
Battery Energy Storage (BES), Compressed Air Energy Storage (CAES),
Superconducting Magnetic Energy storage (SMES), Electrochemical Capacitor Energy
Storage (ECES), and Flywheel Energy Storage (FES).

Modeling ESSs is a complex and time-consuming task due to the number
of different technologies that are currently available and that are
expected to be developed in the future.  While there are several
studies aimed to define the economic viability and the effect on
electricity markets of ESSs (see for example~\cite{beaudin:10}
and~\cite{Awad:14}), there is still no commonly-accepted simple yet
accurate general model of ESSs for voltage and angle stability
studies.  This paper presents a generalized model of ESSs to simplify,
without giving up accuracy, the simulation of different storage
technologies.  The objective is to provide a balanced, fundamental
frequency model that can be defined through a reduced and fixed set of
parameters and that can be readily implemented in power system
simulators for voltage and angle stability analysis.


Studies that focus on ESS dynamic models aimed to
transient stability analysis abound in the literature.
The following are relevant works.  A general taxonomy for ESS dynamic
analysis is presented in \cite{xie:11}.  A detailed 
transient stability model of ECES can be found
in~\cite{intham:13}, while the model of SMES is given
in~\cite{IEEEtaskforce:06}.  Studies of the dynamic performance of a
SMES system coupled to a wind power plant throughout a voltage source
converter (VSC) are presented in~\cite{ali:10} and~\cite{milano:13},
whereas information regarding the installation of a large-scale
  SMES at the Center of Advanced Power Systems at Florida State
  University can be found in~\cite{luongo:03}.  In
\cite{Vongmanee:09}, the authors propose the model of a small-size
CAES based on a polytropic process of air compression and expansion
and in \cite{tremblay:07, shepherd:65} BES dynamic models are proposed
and tested.  Finally, transient stability models of FES can be found
in~\cite{samineni:06} and~\cite{li:06}.

Several simplified ESS models have been also proposed.  Among these, we
cite~\cite{pal:00, wu:12, sui:14, singh:13, fang:14}.  The main
feature that the models proposed in the references above have in
common is that the ESS dynamics are represented considering only
active and reactive power controllers.  However, the dynamics of the
ESS itself are not preserved.  These models are also generally
loss-less as dc and VSC circuitry is neglected.



In this paper, we use the balanced, fundamental frequency model of the
VSC that is proposed in~\cite{Chauduri:14, belmans:14, chauduri:11, Cole:2010, acha:13},
which includes dc circuit and phase-locked loop (PLL) dynamics as well as an 
average $dq$-axis model of the converter and an equivalent model for switching losses.

The model proposed is based on the observation that most ESSs
connected to transmission and distribution grids share a common
structure, i.e., are coupled to the ac network through a VSC device,
present a dc-link and then include another converter (either a dc/dc
or a ac/dc device) to connect the main energy storage device to the dc
link.  Moreover, all storage systems necessarily imply potential and
flow quantities (see Table \ref{table:energies}), whose dynamics
characterize the transient response of the ESS.  In this paper, we
show that, to properly capture the dynamic response of the ESS, it is
important to preserve such dynamics along with those of the VSC converter 
and its controllers.  Controller hard limits, whose relevance has been
discussed in~\cite{fang:14} and \cite{ortega:15}, are also considered.

\begin{table}[t]
  \centering
  \caption{Examples of energy storage technologies.}
  \begin{tabular}{| l | l | l | l |}
    \hline
    \multicolumn{1}{|c|}{\bf{Types of}} & 
    \multicolumn{1}{c|}{\bf{Potential Var.}} & 
    \multicolumn{1}{c|}{\bf{Flow Var.}} & 
    \multicolumn{1}{c|}{\bf{Device}} \\ 
    \multicolumn{1}{|c|}{\bf{Storable Energy}} & 
    \multicolumn{1}{c|}{} & 
    \multicolumn{1}{c|}{} & 
    \multicolumn{1}{c|}{}\\ 
    \hline
    \multicolumn{1}{|l|}{Magnetic} & 
    \multicolumn{1}{l|}{Magneto Motive} & 
    \multicolumn{1}{l|}{Flux} & 
    \multicolumn{1}{l|}{SMES} \\ 
    \multicolumn{1}{|l|}{} & 
    \multicolumn{1}{l|}{Force} & 
    \multicolumn{1}{l|}{} & 
    \multicolumn{1}{l|}{}\\ \hline
    Fluid & Pressure & Mass Flow & CAES \\ \hline
    Electrostatic & Electric Potential & Electric Current & ECES \\ \hline
    \multicolumn{1}{|l|}{Electrochemical} & 
    \multicolumn{1}{l|}{Electrochemical} & 
    \multicolumn{1}{l|}{Molar Flow} & 
    \multicolumn{1}{l|}{BES} \\ 
    \multicolumn{1}{|l|}{} & 
    \multicolumn{1}{l|}{Potential} & 
    \multicolumn{1}{l|}{Rate} & 
    \multicolumn{1}{l|}{}\\ 
    \hline
    Rotational & Angular Velocity & Torque & FES\\ 
    \hline
  \end{tabular}
  \label{table:energies}\vspace{-2.5mm}
\end{table} 

In summary, the paper provides the following contributions:
\begin{enumerate}
\item A simple yet accurate generalized storage device model that
  consists of a given set of parameters and linear differential
  algebraic equations (DAEs). This model is characterized by a
  given set of DAEs whose structure and dynamic order is independent
  from the technology used for the ESS. 
\item A balanced, fundamental frequency model of the typical devices
  that compose an ESS, namely, the VSC, the dc link and their
  regulators.  This model is obtained based on a careful
  selection of well-assessed models for transient stability
  analysis. 
\item A comprehensive validation of the dynamic response of the
  proposed ESS model versus detailed transient stability
  models representing specific technologies as well as
  simplified models that include only active and reactive power
  controllers.
\end{enumerate}



The paper is organized as follows.  Section \ref{sec:overview}
outlines the structure of a generic VSC-based ESS.  The schemes of the
regulators that are common to all ESSs, as well as the active and
reactive controllers used for simplified ESS models are also presented
in this section.  Section \ref{sec:genmodel} describes the hypotheses,
the structure and the formulation of the proposed generalized model of
ESS, as well as how this model is able to describe the behavior of a
variety of ESS technologies.  Section \ref{sec:case} compares the
dynamic response of the proposed model with detailed and simplified
models of ESSs for transient stability
analysis. All simulations are based on the WSCC 9-bus
test system.
Finally, section \ref{sec:conclu} draws conclusions and outlines
future work directions.


\section{ESS Controls and Simplified ESS Model}
\label{sec:overview}

This section presents the elements and controllers that regulate the
dynamic response of the ESS.  In this section, we consider exclusively
the elements of the ESS model that are independent from the storage
technology.  We also present the simplified ESS model that is used for
comparison in the case study presented in Section \ref{sec:case}.
\vspace{-2.5mm}
\subsection{Detailed ESS scheme for Transient Stability Analysis}
\label{subsec:ESSscheme}

\begin{figure}[h!]
  \begin{center}
    \psfrag{ESS}[][]{\LARGE{Storage}}
    \psfrag{Grid}[][]{\LARGE{Grid}}
    \psfrag{Storage Control}[][]{\Large{Storage Control}}
    \psfrag{VSC Control}[][]{\Large{VSC Control}}
    \psfrag{vac}{\LARGE{$v_{\rm ac}$}}
    \psfrag{vacref}{\LARGE{$v_{\rm ac}^{\rm ref}$}}
    \psfrag{vdc}{\LARGE{$v_{\rm dc}$}}
    \psfrag{idc}{\LARGE{$i_{\rm dc}$}}
    \psfrag{iac}{\LARGE{$i_{\rm ac}$}}
    \psfrag{vdcref}{\LARGE{$v_{\rm dc}^{\rm ref}$}}
    \psfrag{+}[][]{\Large{$+$}}
    \psfrag{-}[][]{\Large{$-$}}
    \psfrag{w}{\LARGE{$\textit{w} $}}
    \psfrag{wref}{\LARGE{$\textit{w}^{\rm ref}$}}
    \psfrag{u}{\LARGE{$u$}}
    \psfrag{uref}{\LARGE{$u^{\rm ref}$}}
    \psfrag{a}{\LARGE{$m_{\rm d}$}}
    \psfrag{b}{\LARGE{$m_{\rm q}$}}
    \resizebox{0.95\linewidth}{!}{\includegraphics{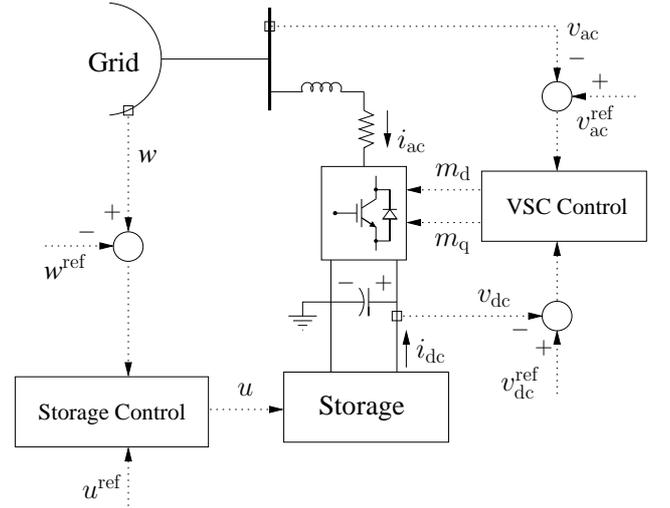}}
    \caption{Scheme of the ESS connected to a grid.}
    \label{ess+vsc}
  \end{center}
  \vspace{-0.2cm}
\end{figure}

Figure \ref{ess+vsc} shows the overall structure of an ESS connected
to a grid through a VSC.  The objective of
the ESS is to control a measured quantity {\it w}, e.g., the frequency
of the Center of Inertia (COI) or the power flowing through a
transmission line.  The elements shown in Fig.~\ref{ess+vsc}
are common to all VSC-based ESSs.  The models
of these elements, except for the storage device,
are presented in the remainder of this section.

\begin{figure}[t!]
  \begin{center}
    \psfrag{+}{\footnotesize{$+$}}
    \psfrag{-}{\footnotesize{$-$}}
    \psfrag{v}{\small{$v_{\rm dc}$}}
    \psfrag{i}{\small{$i_{\rm dc}$}}
    \psfrag{C}{\small{$C_{\rm dc}$}}
    \psfrag{L}{\small{$L_{\rm ac}$}}
    \psfrag{R}{\small{$R_{\rm ac}$}}
    \psfrag{G}{\small{$G_{\rm sw} (i_{\rm dc})$}}
    \psfrag{p}{\small{$p_{\rm ac} + jq_{\rm ac}$}}
    \psfrag{vac}{\small{$v_{\rm ac}  \angle \theta_{\rm ac}$}}
    \psfrag{vt}{\small{$v_{\rm t}  \angle \theta_{\rm t}$}}
    \psfrag{vdc}{\small{$\sqrt{\frac{3}{8}} a_{\rm m} v_{\rm dc}$}}
    \resizebox{\linewidth}{!}{\includegraphics[width=8cm]{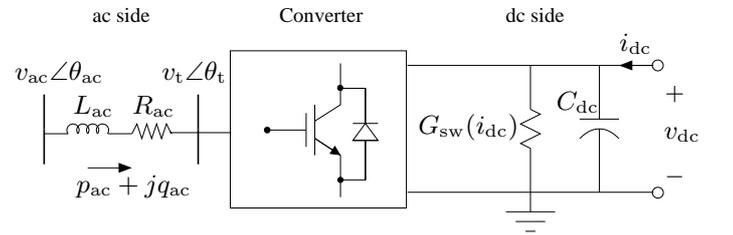}}
    \caption{VSC scheme.}
    \label{vsc}
  \end{center}
  \vspace{-0.2cm}
\end{figure}

Figure \ref{vsc} illustrates the VSC 
scheme~\cite{Chauduri:14, belmans:14, chauduri:11}.
The usual configuration includes a transformer, a bi-directional
converter and a condenser.  The transformer provides galvanic
insulation, whereas the condenser maintains the voltage level at the
dc side of the converter.  The dc voltage is converted to ac voltage
waveform via the use of power electronic converters that are controlled 
using appropriate control logic.
\begin{figure}[t!]
  \begin{center}
    \psfrag{+}{\footnotesize{$+$}}
    \psfrag{-}{\footnotesize{$-$}}
    \psfrag{idref}{\footnotesize{$i_{\rm ac,d}^{\rm ref}$}}
    \psfrag{iqref}{\footnotesize{$i_{\rm ac,q}^{\rm ref}$}}
    \psfrag{id}{\footnotesize{$i_{\rm ac,d}$}}
    \psfrag{iq}{\footnotesize{$i_{\rm ac,q}$}}
    \psfrag{Ki}{\footnotesize{$K_{\rm I} (s)$}}
    \psfrag{wliq}{\scriptsize{$\omega_{\rm ac} L_{\rm ac} i_{\rm ac, q}$}}
    \psfrag{wlid}{\scriptsize{$\omega_{\rm ac} L_{\rm ac} i_{\rm ac, d}$}}
    \psfrag{vacd}{\footnotesize{$v_{\rm ac,d}$}} 
    \psfrag{vacq}{\footnotesize{$v_{\rm ac,q}$}}
    \psfrag{2vdc}{\large{$\frac{2}{v_{\rm dc}}$}}
    \psfrag{vdc2}{\large{$\frac{v_{\rm dc}}{2}$}}
    \psfrag{md}{\footnotesize{$m_{\rm d}$}}
    \psfrag{mq}{\footnotesize{$m_{\rm q}$}}
    \psfrag{vtd}{\scriptsize{$v_{\rm t,d}$}}
    \psfrag{vtq}{\scriptsize{$v_{\rm t,q}$}}
    \psfrag{w}{\footnotesize{$\omega_{\rm ac}$}}
    \psfrag{1}{\footnotesize{$1$}}
    \psfrag{RsL}{\scriptsize{$R_{\rm ac}$+$s L_{\rm ac}$}}    
    \psfrag{L}{\footnotesize{$L_{\rm ac}$}}    
    \psfrag{inner control}{\scriptsize{Inner Current Control Loop}}        
    \psfrag{converter}{\scriptsize{Converter}} 
    \resizebox{0.95\linewidth}{!}{\hspace*{-0.2cm}\includegraphics[width=8cm]{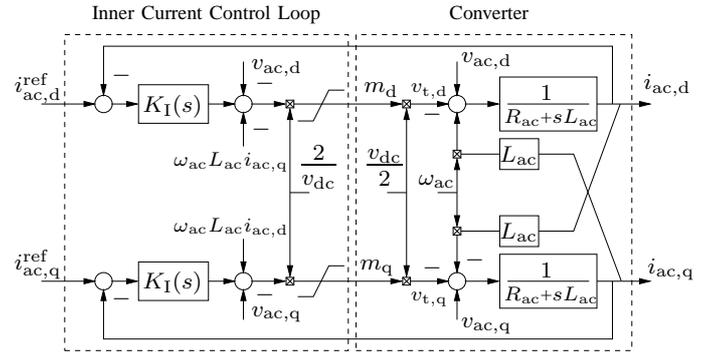}}
    \caption{Block diagram of the inner current control and the converter in the $dq$-frame.}
    \label{vsccoupling}
  \end{center}
  \vspace{-0.4cm}
\end{figure}

Figure~\ref{vsccoupling} shows the block diagram representation of both the converter
and the inner current control loop of the VSC. 
After transforming the three-phase equations to a rotating $dq$-frame, 
the dynamics of the ac side of the VSC depicted in Fig.~\ref{vsc} become:

\begin{equation}
  \begin{array}{lll}
		R_{\rm ac} i_{\rm ac,d} + L_{\rm ac} \dfrac{d i_{\rm ac,d}}{dt} = 
		\hspace*{0.24cm}\omega_{\rm ac} L_{\rm ac} i_{\rm ac, q} + v_{\rm ac, d} \vspace*{0.15cm}  - v_{\rm t,d}\\ 
		R_{\rm ac} i_{\rm ac,q} + L_{\rm ac} \dfrac{d i_{\rm ac,q}}{dt} = 
	 -\omega_{\rm ac} L_{\rm ac} i_{\rm ac, d} + v_{\rm ac, q} - v_{\rm t,q}
  \end{array}
  \label{acvsc}
\end{equation}
where $R_{\rm ac} + jL_{\rm ac}$ is the aggregated impedance of the converter reactance
and transformer; $v_{\rm t,d} = m_{\rm d} v_{\rm dc}/2$;
$v_{\rm t,q} = m_{\rm q} v_{\rm dc}/2$; and $\omega_{\rm ac}$ is the frequency of the grid 
voltage, $\bar{v}_{\rm ac}$. 

\begin{figure}[h!]
  \begin{center}
    \psfrag{-}{\small{$-$}}
    \psfrag{ti}{\small{$\theta_{\rm ac}$}}
    \psfrag{to}{\small{$\theta_{dq}$}}
    \psfrag{K1}{\small{$K_{\rm PD}$}}
    \psfrag{sa}{\small{$1 + sT_{\rm 1,LF}$}}
    \psfrag{sb}{\small{$sT_{\rm 2,LF}$}}
    \psfrag{K2}{\small{$K_{\rm VCO}$}}
    \psfrag{s}{\small{$s$}}
    \resizebox{0.95\linewidth}{!}{\hspace*{-0.2cm}\includegraphics[width=8cm]{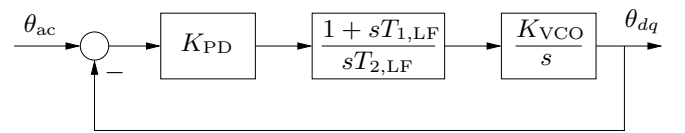}}
    \caption{PLL scheme.}
    \label{pll}
  \end{center}
  \vspace{-0.4cm}
\end{figure}

It is possible to express dc signals in a $dq$-reference frame, with the aim of
designing suitable controllers, provided that the $dq$-reference frame rotates
with the same speed as the grid voltage phasor, which may be achieved, for
example, through the use of a phase-locked loop (PLL)~\cite{Cole:2010}.
The PLL is a control system that forces the angle of the rotating $dq$-frame, 
$\theta_{dq}$, to track the angle $\theta_{\rm ac}$. The
PLL is typically composed of a phase detector (PD), a loop filter (LF), and
a voltage controlled oscillator (VCO), and its scheme is depicted in Fig.~\ref{pll}.

The grid voltage $\bar{v}_{\rm ac}$ can, thus, be 
expressed in the $dq$-frame as follows:
\begin{equation}
	v_{\rm ac,d} + jv_{\rm ac,q} = v_{\rm ac} (\cos(\theta_{\rm ac} - \theta_{dq}) + 
																 j\sin(\theta_{\rm ac} - \theta_{dq}))												
  \label{vacdq}
\end{equation}

The controller $K_{\rm I} (s)$ of Fig.~\ref{vsccoupling} is
designed as a PI control, as follows~\cite{Chauduri:14}:

\begin{equation}
	K_{\rm I} (s) = \dfrac{R_{\rm ac} + s L_{\rm ac}}{s T_{\rm I}}							
  \label{kIvsc}
\end{equation}
where $T_{\rm I}$ is the time-constant of the closed-loop step response.

The power balance between the
dc and the ac sides of the converter is imposed by:

\begin{equation}
	p_{\rm ac} + v_{\rm dc} i_{\rm dc} - p_{\rm loss} - 
		\dfrac{1}{2} C_{\rm dc} \dfrac{d (v_{\rm dc}^2)}{dt} = 0
  \label{balancevsc}
\end{equation}
where $p_{\rm ac} = \dfrac{3}{2} (v_{\rm ac,d} i_{\rm ac,d} + v_{\rm ac,q} i_{\rm ac,q}	)$; 
$\dfrac{1}{2} C_{\rm dc} \dfrac{d (v_{\rm dc}^2)}{dt}$ are the energy variations in
the capacitor;
and $p_{\rm loss} = \dfrac{3}{2} R_{\rm ac} i_{\rm ac}^2 + G_{\rm sw} (i_{\rm dc})v_{\rm dc}^2$
are the circuit and switching losses of the converter, respectively,
with $i_{\rm ac}^2 = i_{\rm ac,d}^2 + i_{\rm ac,q}^2$; 
and $G_{\rm sw} (i_{\rm dc})$ is obtained 
from a given constant conductance, $G_0$, and the quadratic ratio of the actual 
current to the nominal one, as follows~\cite{acha:13}: 
\begin{equation}
  G_{\rm sw} (i_{\rm dc}) = G_0 \left(\frac{i_{\rm dc}}{i_{\rm dc}^{\rm nom}} \right)^2
  \label{VSClosses}
\end{equation}

\begin{figure}[h!]
  \begin{center}
    \psfrag{vac}{$v_{\rm ac}$}%
    \psfrag{vdc}{$v_{\rm dc}$}
    \psfrag{Kmdc}{$K_{\rm mdc}$}
    \psfrag{Tmdc}{$1 + s T_{\rm mdc}$}
    \psfrag{vmdc}{$v_{\rm mdc}$}
    \psfrag{vdcref}{$v_{\rm dc}^{\rm ref}$}
    \psfrag{+}{\tiny{$+$}}    
    \psfrag{-}{\tiny{$-$}}
    \psfrag{PI}{PI control}    
    \psfrag{Kp}{$K_{\rm d}$}
    \psfrag{Ki}{$K_{\rm i, d}$}
    \psfrag{xd}{$x_{\rm d}$}
    \psfrag{s}{$s$}
    \psfrag{idmax}{$i_{\rm d}^{\rm ref, max}$}
    \psfrag{idmin}{$i_{\rm d}^{\rm ref, min}$}
    \psfrag{idref}{$i_{\rm d}^{\rm ref}$}
    \psfrag{Kmac}{$K_{\rm mac}$}
    \psfrag{Tmac}{$1 + s T_{\rm mac}$}
    \psfrag{vmac}{$v_{\rm mac}$}
    \psfrag{vacref}{$v_{\rm ac}^{\rm ref}$}
    \psfrag{+}{\tiny{$+$}}    
    \psfrag{-}{\tiny{$-$}}
    \psfrag{leadlag}{lead/lag control}    
    \psfrag{K}{$K_{\rm q}$}
    \psfrag{KdT2}{$K_{\rm d,q} + s T_2$}
    \psfrag{xq}{$x_{\rm q}$}
    \psfrag{T1}{$1 + s T_1$}
    \psfrag{iqmax}{$i_{\rm q}^{\rm ref, max}$}
    \psfrag{iqmin}{$i_{\rm q}^{\rm ref, min}$}
    \psfrag{iqref}{$i_{\rm q}^{\rm ref}$}
    \resizebox{0.95\linewidth}{!}{\includegraphics[width=9cm]{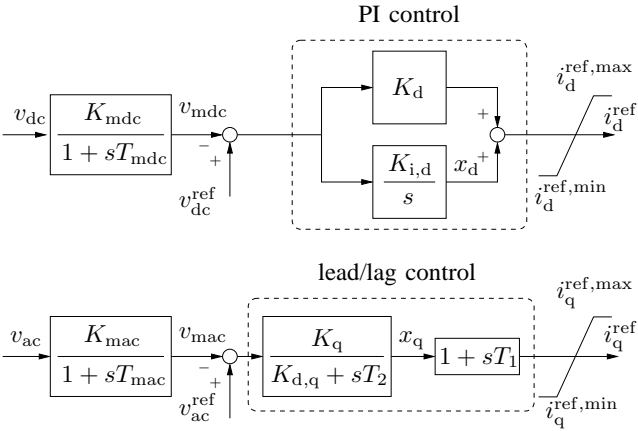}}
    \caption{Outer dc and ac voltage controllers.}
    \label{vscoutercontrol}
  \end{center}
\vspace{-0.4cm}
\end{figure}

Finally, the outer controllers used to regulate the dc and ac voltages of the VSC
are depicted in Fig.~\ref{vscoutercontrol}.  

\begin{figure}[t!]
  \begin{center}
    \psfrag{w}{\huge $\textit{w}$}
    \psfrag{wref}{\huge $\textit{w}^{\rm ref}$}
    \psfrag{E}[][]{\LARGE $E$}
    \psfrag{du}[][]{\huge $|\Delta u|$}
    \psfrag{DB}[][]{\huge Dead-band}
    \psfrag{Ulimit}[][]{\huge Storage Input Limiter}
    \psfrag{1}[][]{\huge $1$}
    \psfrag{1 }{\huge $1$}
    \psfrag{0}{\huge $0$}
    \psfrag{Tf}[][]{\huge$1$+$s T_{\rm f}$}
    \psfrag{xdp}{\huge$x_{\rm f}$}
    \psfrag{+}{}    
    \psfrag{-}{{\large $-$}}
    \psfrag{PI}[][]{\huge PI control}    
    \psfrag{Kp}[][]{\huge$K_{{\rm p}u}$}
    \psfrag{Ki}[][]{\huge$K_{{\rm i}u}$}
    \psfrag{xi}{\huge$x_u$}
    \psfrag{Hd}[][]{\huge$s$}
    \psfrag{umax}{\huge$u^{\rm max}$}
    \psfrag{umin}{\huge$u^{\rm min}$}
    \psfrag{u}{\huge$u$}
    \psfrag{uhat}{\huge$\hat{u}$}
    \psfrag{uref}{\huge$u_{\rm ref}$}
    \resizebox{\linewidth}{!}{\includegraphics{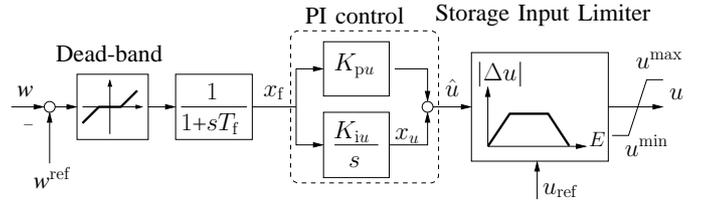}}
    \caption{Storage control scheme.}
    \label{inputsignalcontrol}
  \end{center}
\vspace{-0.4cm}
\end{figure}

The charge/discharge process of the storage device is regu\-lated by
the storage control (see Fig.~\ref{inputsignalcontrol}).  The input
signal of the control is the error between the actual value of a
measured quantity of the system, say $\textit{w}$, and a reference
value ($\textit{w}^{\rm ref}$). If $\textit{w} = \textit{w}^{\rm
  ref}$, the storage device is inactive and its energy is kept
constant.  For $\textit{w} \ne \textit{w}^{\rm ref}$, the storage
device injects active power into the ac bus through the VSC (discharge
process) or absorbs power from the ac bus (charge process). 
The scheme shown in Fig.~\ref{inputsignalcontrol} also includes a block
  referred to as Storage Input Limiter\cite{ortega:15}. The
  purpose of this limiter is to smooth the transients that derive from
  energy saturations of the storage device. 
  This block takes the actual
  value of the energy stored in the device, $E$, and defines the value
  of the controlled input variable of the storage device, $u$, as
  follows:
\begin{equation}
  u = \left\{ \begin{array}{l}
      \displaystyle \frac{E - E_{\rm min}}{E_{\rm min}^{\rm thr} - E_{\rm min}} \Delta u + u_{\rm ref} 
      \vspace*{0.2cm}
      \\ \vspace*{0.2cm}
      \quad \quad \text{if} \quad E_{\rm min} \leq E \leq E_{\rm min}^{\rm thr}
      \quad \text{and} \quad \Delta u > 0 
      \\ \vspace*{0.2cm}
      \displaystyle \frac{E_{\rm max} - E}{E_{\rm max} - E_{\rm max}^{\rm thr}} \Delta u + u_{\rm ref} 
      \\	 \vspace*{0.2cm} 
      \quad \quad \text{if} \quad E_{\rm max}^{\rm thr} \leq E \leq E_{\rm max}
      \quad \text{and} \quad \Delta u < 0\\     	
      \displaystyle \hat{u} \quad \text{otherwise}\\
    \end{array} \right.
  \label{outputlimiter}
\end{equation}
where $u_{\rm ref}$ is the value of $u$ such that the storage device
is disabled; $\Delta u = \hat{u} - u_{\rm ref}$; $E_{\rm min}$ and
$E_{\rm max}$ are the minimum and maximum storable energy in the ESS,
respectively; and $E_{\rm min}^{\rm thr}$ and $E_{\rm max}^{\rm thr}$
are the given minimum and maximum energy thresholds, respectively. 

A dead-band block is also included in Fig.~\ref{inputsignalcontrol}. 
This block is designed to reduce the sensitivity of the Storage
Control to small changes of the measured signal $\textit{w}$ with
respect to the reference $\textit{w}^{\rm ref}$.  The purpose of the
dead-band is to reduce the number of charge/discharge operations of
the ESS \cite{milano:13}. 

\subsection{Simplified ESS scheme}
\label{subsec:simpleESS}

\begin{figure}[h!]
  \begin{center}
    \psfrag{w}{$\textit{w}$}
    \psfrag{wref}{$\textit{w}^{\rm ref}$}
    \psfrag{1}{$1$}
    \psfrag{Tfp}{$1 + s T_{\rm f \it{P}}$}
    \psfrag{xfp}{$x_{\rm f\it{P}}$}
    \psfrag{+}{\tiny{$+$}}    
    \psfrag{-}{\tiny{$-$}}
    \psfrag{PI}{PI control}    
    \psfrag{Kp}[][]{$K_{\rm p \it{P}}$}
    \psfrag{Ki}[][]{$K_{\rm i \it{P}}$}
    \psfrag{xp}{$x_P$}
    \psfrag{Hd}[][]{$s$}
    \psfrag{Tp}{$1 + s T_{P ,\rm ESS}$}
    \psfrag{P}{$P_{\rm ESS}$}
    \psfrag{Pmax}{$P_{\rm ESS}^{\max}$}
    \psfrag{Pmin}{$P_{\rm ESS}^{\min}$}
    \psfrag{v}{$v_{\rm ac}$}
    \psfrag{vref}{$v_{\rm ac}^{\rm ref}$}
    \psfrag{Tfq}{$1 + s T_{\rm f \it{Q}}$}    
    \psfrag{xfq}{$x_{\rm{f}\it{Q}}$}
    \psfrag{leadlag}{lead/lag control}    
    \psfrag{Kq}{$K_Q$}
    \psfrag{T2}{$1 + s T_{Q2}$}
    \psfrag{T1}{$1 + s T_{Q1}$}    
    \psfrag{Tq}{$1 + s T_{Q \rm ,ESS}$}
    \psfrag{Q}{$Q_{\rm ESS}$}
    \psfrag{Qmax}{$Q_{\rm ESS}^{\max}$}
    \psfrag{Qmin}{$Q_{\rm ESS}^{\min}$}
    \resizebox{0.85\linewidth}{!}{\includegraphics[width=9cm]{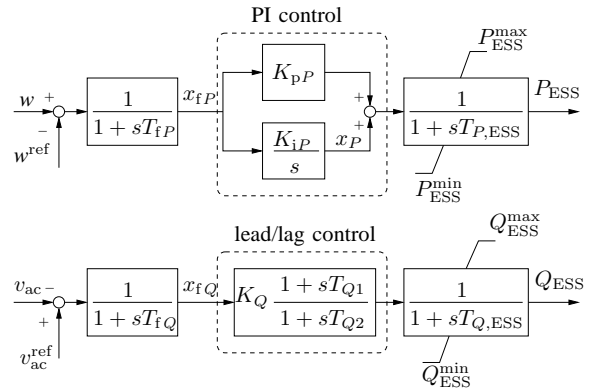}}
    \caption{Scheme of a simplified ESS model.}
    \label{simplescheme}
  \end{center}
\vspace{-0.4cm}
\end{figure}

As discussed in Section \ref{sec:intro}, several simplified models of
ESSs have been proposed \cite{pal:00, wu:12, sui:14, singh:13,
  fang:14}.  For the sake of comparison, in this paper, we consider
the control scheme proposed in \cite{pal:00} (see
Fig.~\ref{simplescheme}).  The input signal $\textit{w}$ is regulated
through the active power while the voltage at the point of connection
with the grid is regulated through the ESS reactive power.  The
physical behavior of the storage system is synthesized by the two lag
blocks with time constants $T_{P, \rm ESS}$ and $T_{Q, \rm ESS}$.

\section{Proposed Generalized Model of Energy Storage Devices}
\label{sec:genmodel}

The generalized model proposed in this paper is based on the
linearization and dynamic reduction of the set of equations that
describes the storage device.  Let assume that the dynamic behavior of
the ESS can be described by a set of nonlinear DAE as
follows:
\begin{alignat}{3}
  \nonumber \boldsymbol{\Gamma} \dot{\boldsymbol{\xi}}(t) &= 
  \boldsymbol{\psi}(\boldsymbol{\xi}(t),\boldsymbol{\nu}(t))\\
  \boldsymbol{\varphi}(t) &=
  \boldsymbol{\eta}(\boldsymbol{\xi}(t),\boldsymbol{\nu}(t))
  \label{DAE}
\end{alignat}
where $\boldsymbol{\xi}(t)$ is the state vector ($\boldsymbol{\xi} \in
\mathbb{R}^n$); $\boldsymbol{\nu}(t)$ is the input vector
($\boldsymbol{\nu} \in \mathbb{R}^m$); $\boldsymbol{\varphi}(t)$ is
the output vector ($\boldsymbol{\varphi} \in \mathbb{R}^p$);
$\boldsymbol{\psi} : \mathbb{R}^{n+m} \rightarrow \mathbb{R}^n$ are
the differential equations; $\boldsymbol{\eta} : \mathbb{R}^{n+m}
\rightarrow \mathbb{R}^p$ are the output equations; and
$\boldsymbol{\Gamma}$ is a diagonal matrix of dimensions $n \times n$
such that \cite{fabozzi:14}:
\begin{list}{}{}
\item $\boldsymbol{\Gamma}_{ii} = 1$ if the $i$-th equation of
  $\boldsymbol{\psi}$ is differential;
\item $\boldsymbol{\Gamma}_{ii} = 0$ if the $i$-th equation of
  $\boldsymbol{\psi}$ is algebraic.
\end{list}	

The first step towards the definition of the generalized model of the
ESS is to linearize the system around the equilibrium point
($\boldsymbol{\xi}_0, \boldsymbol{\varphi}_0,
\boldsymbol{\nu}_0$). Thus, the expression for a linear time-invariant
dynamical system is obtained:
\begin{alignat}{3}
  \nonumber \boldsymbol{\Gamma} \dot{\boldsymbol{\xi}}(t) &= 
  \mathbf{A} \boldsymbol{\xi}(t) + \mathbf{B} \boldsymbol{\nu}(t) + \mathbf{K}_{\xi}\\
  \boldsymbol{\varphi}(t) &= \mathbf{C} \boldsymbol{\xi}(t) +
  \mathbf{D} \boldsymbol{\nu}(t) + \mathbf{K}_{\varphi}
  \label{LTIDS}
\end{alignat}
where $\mathbf{A}$ is the state matrix $\left( {\rm dim}[\mathbf{A}] =
  n \times n \right)$; $\mathbf{B}$ is the input matrix $\left( {\rm
    dim}[\mathbf{B}] = n \times m \right)$; $\mathbf{C}$ is the output
matrix $\left( {\rm dim}[\mathbf{C}] = p \times n \right)$;
$\mathbf{D}$ is the feedthrough matrix $\left( {\rm dim}[\mathbf{D}] =
  p \times m \right)$; and $\mathbf{K}_{\xi} \in \mathbb{R}^n$ and
$\mathbf{K}_{\varphi} \in \mathbb{R}^p$ account for the values of the
variables at the equilibrium point.  Equation (\ref{LTIDS}) is written
for $\bfg \xi$, $\bfg \nu$ and $\bfg \varphi$, not the incremental
values $\Delta \bfg \xi$, $\Delta \bfg \nu$, and $\Delta \bfg
\varphi$.

Then, the state vector $\bfg \xi$ is split into two types of
variables: the potential and flow variables related to the energy
stored in the ESS, $\bfg x$ (see Table \ref{table:energies}); and all
other state variables, $\bfg z$.  Hence, $\bfg \xi = [\bfg x^T, \;
  \bfg z^T]^T $.  Since all ESSs are coupled to the ac network
through a VSC device, the dc voltage and current of the VSC, $v_{\rm
  dc}$ and $i_{\rm dc}$, are considered as an input and an output in
(\ref{LTIDS}), respectively.  The input vector $\bfg \nu$ is composed
of the output signals of the storage control, $u$, and the dc voltage
of the VSC, $v_{\rm dc}$.  Hence, $\bfg \nu = [u, \; v_{\rm dc}]^{\rm
  T}$.  Finally, the output vector $\boldsymbol{\varphi}$ is the dc
current of the VSC, $i_{\rm dc}$.  Hence, $\boldsymbol{\varphi} =
[i_{\rm dc}]$.  Applying the notation above, equation (\ref{LTIDS}) is
written as follows:
\begin{alignat}{3}
  \nonumber \mathbf{\Gamma}_x \dot{\boldsymbol{x}} &= 
  \mathbf{A}_{xx}\boldsymbol{x} + \mathbf{A}_{xz}\boldsymbol{z} + 
  \mathbf{B}_{xu} u + \mathbf{B}_{xv} v_{\rm dc} + \mathbf{K}_x\\    
  \label{FullGenModel}
  \mathbf{\Gamma}_z \dot{\boldsymbol{z}} &= \mathbf{A}_{zx}\boldsymbol{x} + 
  \mathbf{A}_{zz}\boldsymbol{z} + 
  \mathbf{B}_{zu} u + \mathbf{B}_{zv} v_{\rm dc} + \mathbf{K}_z\\
  \nonumber i_{\rm dc}&= \mathbf{C}_{x}\boldsymbol{x} + 
  \mathbf{C}_{z}\boldsymbol{z} + \mathbf{D}_{u} u + 
  \mathbf{D}_{v} v_{\rm dc} + \mathbf{K}_i
\end{alignat}
where:
\begin{list}{}{}
\item ${\scriptscriptstyle{\begin{bmatrix} \mathbf{A}_{xx} &
        \mathbf{A}_{xz} \\ \mathbf{A}_{zx} &
        \mathbf{A}_{zz} \end{bmatrix}}} = \mathbf{A}$ ;
  \;${\scriptscriptstyle{\begin{bmatrix} \mathbf{B}_{xu} &
        \mathbf{B}_{xv} \\ \mathbf{B}_{zu} &
        \mathbf{B}_{zv} \end{bmatrix}}} = \mathbf{B}$ ; \vspace*{2mm} \item
  $[\mathbf{C}_{x} \; \mathbf{C}_{z}] = \mathbf{C}$ ; \;
  $[\mathbf{D}_{u} \; \mathbf{D}_{v}] = \mathbf{D}$ ; \vspace*{2mm} \item
  $[\mathbf{\Gamma}_{x} \; \mathbf{\Gamma}_{z}]^T =
  \mathbf{\Gamma}$ ; \; $[\mathbf{K}_{x} \; \mathbf{K}_{z}]^T =
  \mathbf{K}_{\xi}$ ; \; $\mathbf{K}_{i} = \mathbf{K}_{\varphi}$
\end{list}	

The dynamic order of (\ref{FullGenModel}) is reduced by assuming that
the transient response of $\boldsymbol{z}$ are much faster than that
of $\boldsymbol{x}$ or immaterial with respect to the overall dynamic
behavior of the ESS.  This assumption is based on the knowledge of
detailed transient stability models and is
duly verified through the simulations presented in Section
\ref{sec:case}.  By neglecting the dynamics of $\boldsymbol{z}$ (i.e.,
$\mathbf{\Gamma}_{z} = \mathbf{0}$), from the second equation of
(\ref{FullGenModel}), we obtain:
\begin{equation}
  \boldsymbol{z} = - \mathbf{A}_{zz}^{-1} \left(
  \mathbf{A}_{zx}\boldsymbol{x} + \mathbf{B}_{zu}u +
  \mathbf{B}_{zv} v_{\rm dc} + \mathbf{K}_{z} \right)
  \label{ClearedZeta}
\end{equation}
Then, substituting (\ref{ClearedZeta}) into the first and third
equations of (\ref{FullGenModel}), one has:
\begin{alignat}{3}
  \nonumber \mathbf{\Gamma}_x \dot{\boldsymbol{x}} &=
  \left(\mathbf{A}_{xx} - \mathbf{A}_{xz}\mathbf{A}_{zz}^{-1}
    \mathbf{A}_{zx}\right)\boldsymbol{x} + \left(\mathbf{B}_{xu} -
    \mathbf{A}_{xz}\mathbf{A}_{zz}^{-1} \mathbf{B}_{zu}\right)u \\
  \nonumber &+ \left(\mathbf{B}_{xv} -
    \mathbf{A}_{xz}\mathbf{A}_{zz}^{-1} \mathbf{B}_{zv}\right)v_{\rm
    dc} +
  \left(\mathbf{K}_{x} - \mathbf{A}_{xz}\mathbf{A}_{zz}^{-1} \mathbf{K}_{z}\right)\\
  \nonumber i_{\rm dc} &= \left(\mathbf{C}_{x} -
    \mathbf{C}_{z}\mathbf{A}_{zz}^{-1}\mathbf{A}_{zx}\right)
  \boldsymbol{x} + \left(\mathbf{D}_{u} - \mathbf{C}_{z}
    \mathbf{A}_{zz}^{-1}\mathbf{B}_{zu}\right) u \\
  &+ \left(\mathbf{D}_{v} -
    \mathbf{C}_{z}\mathbf{A}_{zz}^{-1}\mathbf{B}_{zv}\right) v_{\rm
    dc} + \left(\mathbf{K}_{i} -
    \mathbf{C}_{z}\mathbf{A}_{zz}^{-1}\mathbf{K}_{z}\right)
  \label{w/o_zeta}
\end{alignat}

Rewriting in compact form the matrices in (\ref{w/o_zeta}), the
proposed generalized model of energy storage devices can be written as
follows:
\begin{alignat}{3}
  \nonumber \tilde{\mathbf{\Gamma}}\dot{\boldsymbol{x}} &=&
  \tilde{\mathbf{A}}\boldsymbol{x} + \tilde{\mathbf{B}}_u
  u +
  \tilde{\mathbf{B}}_v v_{\rm dc} + \tilde{\mathbf{K}}_x\\
  i_{\rm dc} &=& \tilde{\mathbf{C}}\boldsymbol{x} +
  \tilde{\mathbf{D}}_u u + \tilde{\mathbf{D}}_v v_{\rm
    dc} + \tilde{\mathbf{K}}_i
  \label{GenModel}	
\end{alignat}
It is important to take into account the state of charge of the
storage device (in particular, to impose energy limits to the
controller shown in Fig.~\ref{inputsignalcontrol}). With this aim, 
the actual stored energy is given by:
\begin{equation}
  E = \sum_{\rm i=1}^{n} \rho_{\rm i} \left(x_{\rm i}^{\beta_{\rm i}} 
    - \chi_{\rm i}^{\beta_{\rm i}} \right) 
  \label{GenEnergy}
\end{equation}
where $\rho_{\rm i}$, $\beta_{\rm i}$ and $\chi_{\rm i}$ are the
proportional coefficient, exponential coefficient and reference
potential value of each variable $x_{\rm i}$, respectively.

In summary, the proposed generalized ESS model is composed of 
the block diagram of the VSC represented in Fig.~\ref{vsccoupling}; 
the controllers depicted in Figs.~\ref{vscoutercontrol} 
and \ref{inputsignalcontrol}; and (\ref{GenModel}), (\ref{GenEnergy}).

For the sake of clarity and example, in the following subsections we
deduce the generalized model for a variety of ESSs technologies.

\vspace{-2mm}
\subsection{Electrochemical Capacitor Energy Storage Model}
\label{subsec:sces}

\begin{figure}[h!]
  \begin{center}
    \psfrag{Vsc}{{\large $v_{\rm sc}$}}
    \psfrag{Gsc}{{\large$G_{\rm sc}$}}
    \psfrag{Lsc}{{\large$L_{\rm sc}$}}
    \psfrag{Rsc}{{\large$R_{\rm sc}$}}
    \psfrag{Isc}{{\large$i_{\rm sc}$}}
    \psfrag{Csc}{{\large$C_{\rm sc}$}}
    \psfrag{DC}{{\large$\rm DC$}}
    \psfrag{Vdc}{{\large$v_{\rm dc}$}}
    \psfrag{Idc}{{\large$i_{\rm dc}$}}
    \psfrag{+}{$+$}
    \psfrag{-}{$-$}
    \psfrag{S}{{\large$S$}}
    \resizebox{0.7\linewidth}{!}{\includegraphics[width=7.5cm]{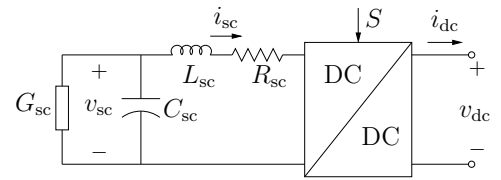}}
    \caption{Scheme of an ECES.}
    \label{sces}
  \end{center}
\vspace{-0.4cm}
\end{figure}

Figure \ref{sces} shows the scheme of an ECES connected to the VSC through a bidirectional dc/dc converter
(buck-boost)~\cite{intham:13}.
In the buck operation mode, 
the energy is delivered from the storage device to the grid, 
while in the boost mode, the energy is stored in the ECES.

The model that describes both the buck and the boost operation modes
is as follows:
\begin{alignat}{3}
  \nonumber & \dot{v}_{\rm sc} &=& -\frac{1}{C_{\rm sc}} \left( i_{\rm sc} 
    + G_{\rm sc} v_{\rm sc} \right) \\
  \label{scesmodel}
  & \dot{i}_{\rm sc} &=& \ \frac{1}{L_{\rm sc}} \left( v_{\rm sc} -
    i_{\rm sc} R_{\rm sc} - v_{\rm dc} S \right) \\
  \nonumber & i_{\rm dc} &=& \ S i_{\rm sc}  
\end{alignat}
where $S$ is the logic state of the switches of the converter. Assu\-ming
average variable values, $S$ is continuous and 
re\-presents the duty cycle of the converter. 

The energy stored in the ECES is as follows: 
\begin{equation}
  E = \frac{1}{2} C_{\rm sc} v_{\rm sc}^2 + \frac{1}{2} L_{\rm sc}
  i_{\rm sc}^2
  \label{scesenergy}
\end{equation}

In (\ref{scesenergy}), the main term is that related to the
capacitance $C_{\rm sc}$, as expected.  However, for completeness, we
include also the energy stored in the inductance $L_{\rm sc}$.

Applying the notation of the proposed general model (\ref{GenModel}-\ref{GenEnergy})
to (\ref{scesmodel}-\ref{scesenergy}), ECES variables and
parameters are:
\begin{alignat}{1}
  \nonumber & \boldsymbol{x} = [v_{\rm sc} \; \; i_{\rm sc}]^T; \;
  \boldsymbol{z} = \emptyset; \; \; u = S; \; \;
  \tilde{\mathbf{\Gamma}} = {\scriptscriptstyle{\begin{bmatrix} 1 & 0
        \\ 0 & 1 \end{bmatrix}}} ; \\ \nonumber & \tilde{\mathbf{A}} =
        {\scriptscriptstyle{\begin{bmatrix} -\frac{G_{\rm sc}}{C_{\rm
                  sc}} & -\frac{1}{C_{\rm sc}} \\ \frac{1}{L_{\rm sc}}
              & -\frac{R_{\rm sc}}{L_{\rm sc}} \end{bmatrix}}}
        ;\ \tilde{\mathbf{B}}_u = \left[ 0 \; \; \frac{v_{\rm
              dc,0}}{L_{\rm sc}} \right]^T; \; \tilde{\mathbf{B}}_v =
        \left[ 0 \; \; \frac{S_{\rm 0}}{L_{\rm sc}} \right]^T;
        \\ \nonumber & \tilde{\mathbf{K}}_x = {\scriptscriptstyle
          {\begin{bmatrix} \frac{1}{C_{\rm sc}} \left( G_{\rm
                sc}v_{\rm sc,0} + i_{\rm sc,0} \right)
              \\ \frac{1}{L_{\rm sc}} \left( -v_{\rm sc,0} + R_{\rm
                sc} i_{\rm sc,0} + 2 v_{\rm dc,0} S_{\rm 0}
              \right) \end{bmatrix}}}; \; \tilde{\mathbf{C}} = [0 \;
          \; S_{\rm 0}] ; \\ \nonumber & \tilde{\mathbf{D}}_u =
           [i_{\rm sc,0}] ;\ \tilde{\mathbf{D}}_v =
           [0];\ \tilde{\mathbf{K}}_i = [i_{\rm dc,0} - 2 S_{\rm 0}
             i_{\rm sc,0}]; \\ \nonumber & \boldsymbol{\rho} = \left[
             \frac{1}{2} C_{\rm sc} \; \; \frac{1}{2} L_{\rm sc}
             \right] ;\ \boldsymbol{\beta} = [2 \; \; 2]
           ;\ \boldsymbol{\chi} = [0 \; \; 0]
\end{alignat}
where $S_{\rm 0}$, $v_{\rm sc,0}$, $v_{\rm dc,0}$, $i_{\rm sc,0}$ and
$i_{\rm dc,0}$ are the values of $S$, $v_{\rm sc}$, $v_{\rm dc}$,
$i_{\rm sc}$ and $i_{\rm dc}$ at the equilibrium point, respectively.

\subsection{Superconducting Magnetic Energy Storage Model}
\label{subsec:smes}
\begin{figure}[htb]
  \begin{center}
   \psfrag{Ic}[c][c]{\Huge $i_{\rm c}$}
    \psfrag{Vc}[c][c]{\Huge $v_{\rm c}$}
    \psfrag{Idc}[c][c]{\Huge $i_{\rm dc}$}
    \psfrag{Vdc}[c][c]{\Huge $v_{\rm dc}$}
    \psfrag{DC}[c][c]{\Huge $\rm DC$}
    \psfrag{S}[c][c]{\Huge $S$}
    \psfrag{+}[c][c]{\LARGE $+$}
    \psfrag{-}[c][c]{\LARGE $-$}
    \resizebox{4cm}{!}{\includegraphics{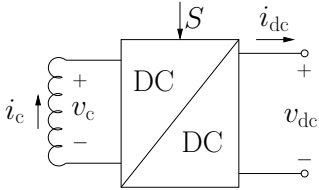}}
    \caption{Scheme of the SMES and the boost converter.}
    \label{fig:SMES}
  \end{center}
\vspace{-0.4cm}
\end{figure}

Figure \ref{fig:SMES} depicts the scheme of a 
SMES~\cite{IEEEtaskforce:06}.  The SMES is connected
in parallel to the VSC throughout a dc/dc converter (boost converter).
The SMES stores magnetic energy and injects it into the network
according to the duty cycle applied to the converter.

The dynamics of the coil and the dc/dc converter are represented by
the circuit equations:
\begin{alignat}{3}
  \nonumber & \dot{i}_{\rm c} &=& -\frac{v_{\rm c}}{L_{\rm c}} \\
  \label{smesmodel}
  &\ 0 &=& \ (1 - 2 S) v_{\rm dc} - v_{\rm c}\\
  \nonumber & i_{\rm dc} &=& \ (1 - 2 S) i_{\rm c}  
\end{alignat}
where all voltages and currents are average values; $L_{\rm c}$ is the
inductance of the superconducting coil; and $S$ is the duty cycle of
the dc/dc converter.

The energy stored in the SMES is given by:
\begin{equation}
  E = \frac{1}{2} L_{\rm c} i_{\rm c}^2
\end{equation}

Therefore, using the notation of the proposed model:
\begin{alignat}{3}
  \nonumber & \boldsymbol{x} = [i_{\rm c} \; \; v_{\rm c}]^T; \;
  \boldsymbol{z} = \emptyset; \; \; u = S; \; \;
  \tilde{\mathbf{\Gamma}} = {\scriptscriptstyle{\begin{bmatrix} 1 & 0
        \\ 0 & 0 \end{bmatrix}}} \\ & \boldsymbol{\rho} =
  \left[\frac{1}{2} L_{\rm c} \; \; 0 \right] ; \; \;
  \boldsymbol{\beta} = \left[ 2 \; \; 1 \right]; \; \;
  \boldsymbol{\chi} = \left[0 \; \; 0 \right]
\end{alignat}

\vspace{-5mm}

\subsection{Compressed Air Energy Storage Model}
\label{subsec:caes}

\begin{figure}[htb]
  \begin{center}
    \psfrag{TANK}[][]{\huge $\rm TANK$}
    \psfrag{COMP}[][]{\huge $\rm COMP$}
    \psfrag{TURB}[][]{\huge $\rm TURB$}
    \psfrag{AC}[][]{\LARGE $\rm AC$}
    \psfrag{DC}[][]{\LARGE $\rm DC$}
    \psfrag{+}[][]{\Large $+$}
    \psfrag{-}[][]{\Large $-$}
    \psfrag{vt}[][]{\Huge $v_{\rm dc}$}
    \psfrag{it}[][]{\Huge $i_{\rm dc, t}$}
    \psfrag{at}[][]{\Huge $\hat{m}_{\rm d,t}$}
    \psfrag{bt}[][]{\Huge $\hat{m}_{\rm q,t}$}
    \psfrag{vc}[][]{\Huge $v_{\rm dc}$}
    \psfrag{ic}[][]{\Huge $i_{\rm dc, c}$}
    \psfrag{ac}[][]{\Huge $\hat{m}_{\rm d,c}$}
    \psfrag{bc}[][]{\Huge $\hat{m}_{\rm q,c}$}
    \resizebox{\linewidth}{!}{\includegraphics{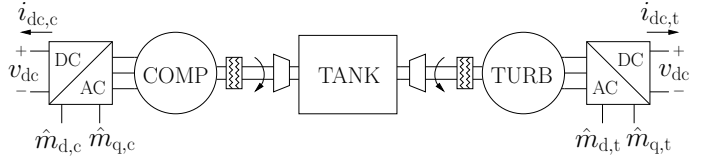}}
    \caption{Scheme of a Compressed Air Energy Storage.}
    \label{fig:caes}
  \end{center}
\vspace{-0.4cm}
\end{figure}

The basic configuration of a CAES is
represented in Fig. \ref{fig:caes}~\cite{Vongmanee:09}.  The air is
injected into the tank by means of a compressor operated by an
asynchronous motor, and extracted from it through a turbine driven by
an asynchronous generator.  Both the compressor and the turbine are
connected in parallel to the dc link of the VSC through ac/dc
converters.

The air is modeled as an ideal gas.  Therefore, the variation
of the pressure inside the tank can be calculated as follows:
\begin{equation}
  \dot{\Pi}_{2}=\frac{\rho R\Theta_{2}}{{\pi_{\rm m} \rm Vol}} Q
  \label{Pressure}
\end{equation}
where $\Pi_{2}$ is the pressure inside the tank; $\rho$ is the air
density; $R$ is the ideal gas constant; $\Theta_{2}$ is the
temperature of the air inside the tank; $\pi_{\rm m}$ is the molecular
weight of air; $\rm Vol$ is the volume of the tank; and $Q$ is the air
flow through the compressor/turbine (all quantities are assumed in
absolute value).


Assuming that both compression and expansion of air are polytropic
processes~\cite{Vongmanee:09}, the behavior of the compressor of the
CAES is described by the following set of nonlinear equations (turbine
equations are similar and are omitted):
\begin{alignat}{3}
  \label{efpower} &0 &=&\ \frac{\gamma}{\gamma-1} \Pi_{1} Q \left[ 
    \left( \frac{\Pi_{2}}{\Pi_{1}} \right)^{\frac{\gamma-1}{\gamma}}
    -1 \right] - P_{\rm ef} \\ 
    \label{Omega} &\dot{\Omega} &=&\ \frac{1}{2H}(T_{\rm el}-T_{\rm m})\\ 
    &0 &=&\ P_{\rm ef} - \eta_{\rm m}P_{\rm m}\\ 
    &0 &=&\ \frac{P_{\rm m}}{\Omega} - T_{\rm m}\\ 
    &0 &=&\ T_{\rm el}\ \Omega - P_{\rm el}\\ 
    &0 &=&\ \frac{\omega_{\rm 1n} - \omega_{\rm 2}}{\omega_{\rm 1n}} - \sigma \\
  \label{OmegaCAES}
  &0 &=&\ n_{\rm p} \Omega - \omega_{\rm 2}
\end{alignat}
where $\gamma$ is the polytropic exponent of the air; $\Pi_1$ is the
atmos\-pheric pressure; $P_{\rm ef},\ P_{\rm m}$ and $P_{\rm el}$ are
the effective, mechanical and electrical powers of the compressor,
respectively; $T_{\rm m}$ and $T_{\rm el}$ are the mechanical and
electrical torques of the motor, respectively; $\Omega$ and
$\omega_{\rm 2}$ are the mechanical and electrical rotor speed of the
compressor, respectively; $\sigma$ is the motor slip; 
and $\eta_{\rm m}$, $n_{\rm p}$, $\omega_{\rm 1n}$, 
$H$ are motor parameters.

For the sake of space, the equations of the electrical machines
  connected to the compressor and turbine are not shown in
  (\ref{efpower})-(\ref{OmegaCAES}).  The model considered for the
  electric machines is a $5^{\rm th}$ order $dq$-model (see, for
  example, Chapter 4 of~\cite{Krause:02}).  Finally, the ac/dc
  converters are controlled in a similar way as the VSC described in
  Section \ref{sec:overview}.

According to (\ref{efpower}) and (\ref{Omega}), the energy stored in
the CAES can be expressed as follows:
\begin{equation}
  E=\frac{\gamma}{\gamma -1}\Pi_1 {\rm Vol}
  \left[\left(\frac{\Pi_{2}}{\Pi_{1}}\right)^{\frac{\gamma-1}{\gamma}}
    -1\right] + H \Omega^2
\end{equation} 

As in Subsection \ref{subsec:sces}, both terms related to the energy
stored in the CAES have been considered, despite the fact that the
term related to the mechanical rotor speed $\Omega$ is expected to be
much smaller than the term related to the pressure $\Pi_2$.

Taking into account that the compressor and the turbine have similar
sizes, we assume that both compressor and turbine electrical machines
have same parameters.  Therefore, we model only one equivalent
electrical machine capable of working, depending on the sign of $Q$,
as a compressor ($Q > 0$) or a turbine ($Q < 0$).

Imposing the proposed generalized model to CAES equations, we obtain:
\begin{alignat}{3}
  \nonumber & \boldsymbol{x} = [\Pi_2 \; \; \Omega]^T; \; u = Q; \; \;
  i_{\rm dc} = i_{\rm dc,c} + i_{\rm dc,t}; \\
  \label{caesgen}
   &   \boldsymbol{z} = [P_{\rm ef} \; \; P_{\rm m} \; \; P_{\rm el} \; \;
    T_{\rm m} \; \; T_{\rm el} \; \; \sigma \; \; \omega_2 \;
    \; \hat{m}_{\rm d} \; \; \hat{m}_{\rm q} \; \; \dots]^T;  \\ 
  \nonumber & \boldsymbol{\rho} =
  \left[\frac{\gamma}{\gamma - 1} \Pi_{\rm 1}^{\gamma ^ {- 1}} {\rm
      Vol} \; \; H \right] ; \; \; \boldsymbol{\beta} = \left[
    \frac{\gamma - 1}{\gamma} \; \; 2 \right]; \; \; \boldsymbol{\chi} =
  \left[\Pi_{\rm 1} \; \; 0 \right]
\end{alignat}
where the dots in the vector $\bfg z$ stand for
electrical machine and VSC variables which are not explicitly defined
in (\ref{efpower})-(\ref{OmegaCAES}).  $\bfg z$ contains $25$ state
and algebraic variables. 

Since we do not model the air valve, the output of the storage
control, and thus the input to the generalized model $u$ is the air
flow, $Q$.  By using the proposed model, the nonlinear set of DAEs
which models the CAES ((\ref{Pressure})-(\ref{OmegaCAES}),
electrical machine equations of the compressor and turbine, and
equations of the ac/dc converters) is reduced to a set of three linear
DAEs (\ref{GenModel}).


\subsection{Battery Energy Storage Model}
\label{subsec:bess}

A commonly-used model to represent the dynamics of a rechargeable battery
cell is the Shepherd model~\cite{shepherd:65}:
\begin{alignat}{3}
  \nonumber & \dot{Q}_{\rm e} \ &=& \ i_{\rm b}/3600 \\
  \label{bessmodel}	
  & \dot{i}_{\rm m} \ &=&\ \frac{i_{\rm b} - i_{\rm m}}{T_{\rm m}}
  \\ \nonumber & 0 \ &=& \ v_{\rm oc} - v_{\rm p}(Q_{\rm e}, i_{\rm
    m}) + v_{\rm e} e ^{-\beta _{\rm e} Q_{\rm e}} - R_{\rm i} i_{\rm
    b} - v_{\rm b}
\end{alignat}
where $Q_{\rm e}$ is the extracted capacity in Ah; $i_{\rm m}$ is the
battery current $i_{\rm b}$ passed through a low-pass filter with time
constant $T_{\rm m}$; $v_{\rm oc}$, $v_{\rm p}$ and $v_{\rm e}$ are
the open-circuit, polarization and exponential voltages, respectively;
$\beta_{\rm e}$ is the exponential zone time constant inverse; $R_{\rm
  i}$ is the internal battery resistance; and $v_{\rm b}$ is the
battery voltage.

The variation of the polarization voltage $v_{\rm p}$ with respect to
$i_{\rm m}$ and $Q_{\rm e}$ is given by:

\begin{equation}
  v_{\rm p}(Q_{\rm e}, i_{\rm m}) = \left\{ \begin{array}{l}
    \displaystyle \frac{R_{\rm p} i_{\rm m} + K_{\rm p} Q_{\rm e}}{\rm
      SOC} \quad \text{if} \quad i_{\rm m} > 0 \; \;
    \text{(discharge)}\\ \displaystyle \frac{R_{\rm p} i_{\rm
        m}}{q_{\rm e} + 0.1} + \frac{K_{\rm p} Q_{\rm e}}{\rm SOC}
    \quad \text{if} \quad i_{\rm m} \leq 0 \; \; \text{(charge)}\\
    \end{array} \right. 
\end{equation}
where $R_{\rm p}$ and $K_{\rm p}$ are the polarization resistant and
polarization constant, respectively; and SOC is the state of charge of
the battery which is defined as:
\begin{equation}
  \text{SOC} = \frac{Q_{\rm n} - Q_{\rm e}}{Q_{\rm n}} = 1 - q_{\rm e}
\end{equation}

The non-linearity of $v_{\rm p}$ implies that depending on the state
of the battery (charge or discharge), two different sets of equations
can be obtained by applying the proposed generalized model.
Therefore, the proposed model has to be able to switch from one set to
another depending on the BES operation.

The connection of the BES to the VSC is similar to the
SMES (see Subsection \ref{subsec:smes}):
\begin{alignat}{3}
  \nonumber & 0 &=& \ (1 - 2S) v_{\rm dc} - n_{\rm s} v_{\rm b} \\
  \label{bessvsc}	
  & i_{\rm dc} \ &=&\  - (1 - 2S) n_{\rm p} i_{\rm b}
\end{alignat}
where $S$ is the duty cycle of the converter; and $n_{\rm p}$ and
$n_{\rm s}$ are the number of parallel and series connected battery
cells, respectively. The dc/dc converter connection used in this
  paper is based on~\cite{esmaili:13}.  Other configurations are also
  possible, but are not considered in this paper.

Imposing the proposed generalized model to BES equations, one has:
\begin{alignat}{3}
  \nonumber &\boldsymbol{x} = [Q_{\rm e} \; \; v_{\rm b}]^T; \;
  \boldsymbol{z} = [i_{\rm b} \; \; i_{\rm m} \; \; v_{\rm p}]^T;
  \; \; u = S; \; \tilde{\mathbf{\Gamma}} =
     {\scriptscriptstyle{\begin{bmatrix} 1 & 0 \\ 0 &
           0 \end{bmatrix}}} ; \\ &\boldsymbol{\rho} =
     \left[-\frac{1}{Q_{\rm n}} \; \; 0 \right]; \; \;
     \boldsymbol{\beta} = [1 \; \; 1]; \; \; \boldsymbol{\chi}
     = [Q_{\rm n} \; \; 0]
\end{alignat}


\section{Case Study}
\label{sec:case}

\begin{figure}[h!]
  \begin{center}
    \psfrag{ESS}[][]{\Large $\rm ESS$}
    \psfrag{G}[][]{\Large $\rm G$}
    \psfrag{1}[][]{\Large $1$}
    \psfrag{2}[][]{\Large $2$}
    \psfrag{3}[][]{\Large $3$}
    \psfrag{4}[][]{\Large $4$}
    \psfrag{5}[][]{\Large $5$}
    \psfrag{6}[][]{\Large $6$}
    \psfrag{7}[][]{\Large $7$}
    \psfrag{8}[][]{\Large $8$}
    \psfrag{9}[][]{\Large $9$}
    \resizebox{\linewidth}{!}{\includegraphics{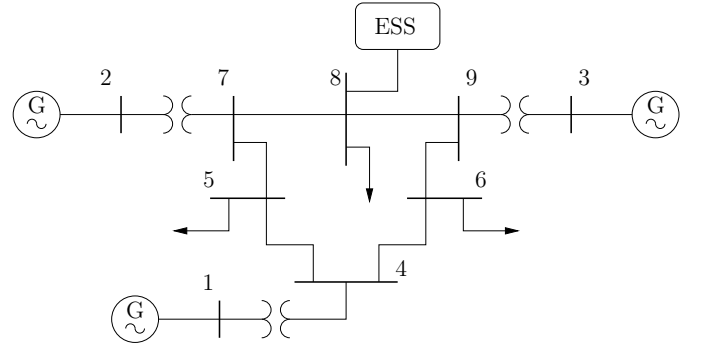}}
    \caption{WSCC 9-bus test system with an ESS device connected to bus 8.}
    \label{9bus}
  \end{center}
\vspace{-0.4cm}
\end{figure}

This section validates the proposed generalized ESS model through time
domain simulations. Three energy storage devices are considered,
namely, SMES, CAES and BES. With this aim, the WSCC 9-bus test system
(see Fig. \ref{9bus}) is used for all simulations.  This benchmark
network consists of three synchronous machines, three transformers,
six transmission lines and three loads.  Primary frequency and voltage
regulators (AVRs) are also included. 
All dynamic data of the WSCC
9-bus system as well as a detailed discussion of its transient
behavior were originally provided in~\cite{Anderson:03}, and 
are publicly on several websites, e.g.,~\cite{web:9bus}.

An ESS is connected to bus 8.  The capacities of the ESSs used in this
case study have been designed according to the power installed in the
system.  This leads to assume large ESSs. 


Different scenarios have been performed in this paper: Subsection
\ref{subsec:smesTDS} shows the response of a SMES connected to the
WSCC system following a three-phase fault (\ref{subsub:smesdeterm}), 
and stochastic variations of the loads (\ref{subsub:smesstoc}).
A similar analysis is carried out in Subsection~\ref{subsec:caesTDS} for a CAES device.
In this case, the contingency is a loss of load (\ref{subsub:caesdeterm}),
and stochastic processes are applied to all load power consumptions (\ref{subsub:caesstoc}).
Finally, a BES and a loss of load are considered in Subsection~\ref{subsec:bessTDS}.
Note that results shown in this section are not hand-picked but, rather, randomly
selected among several hundreds of simulations that have been carried out to check
the validity and accuracy of the proposed generalized ESS model.

All simulations and plots have been obtained using \textsc{Dome}
\cite{Vancouver}.  \textsc{Dome} has been compiled based on Python
2.7.5, CVXOPT 1.1.5, SuiteSparse 4.2.1, and Matplotlib 1.3.0; and has
been executed on a 64-bit Linux Ubuntu 12.04 distribution running on 8
core 3.60 GHz Intel Xeon with 12 GB of RAM.

\vspace{-2mm}

\begin{figure}[t!]
  \centering
  \subfigure{\parbox[t]{.03\linewidth}{\small (a)\hfill}
    {\parbox[c]{\linewidth}{{\label{SMES_COI_noSaturation}
          \resizebox{1.03\linewidth}{!}{\includegraphics{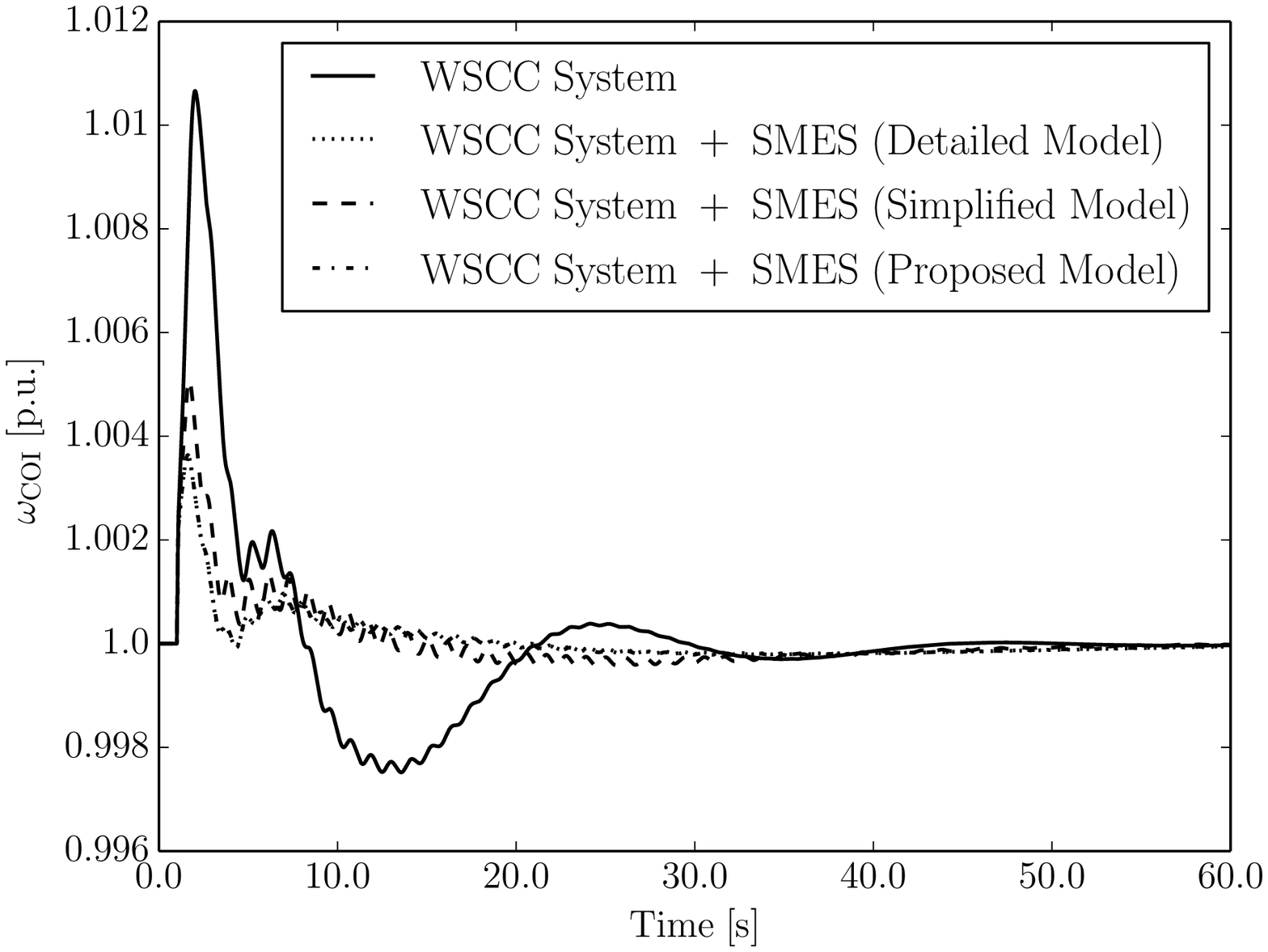}} \vspace{-8mm}}}}}
  \vspace{-2.5mm}\subfigure{\parbox[t]{.032\linewidth}{\small (b)\hfill}
    {\parbox[c]{1.03\linewidth}{\resizebox{\linewidth}{!}{\includegraphics{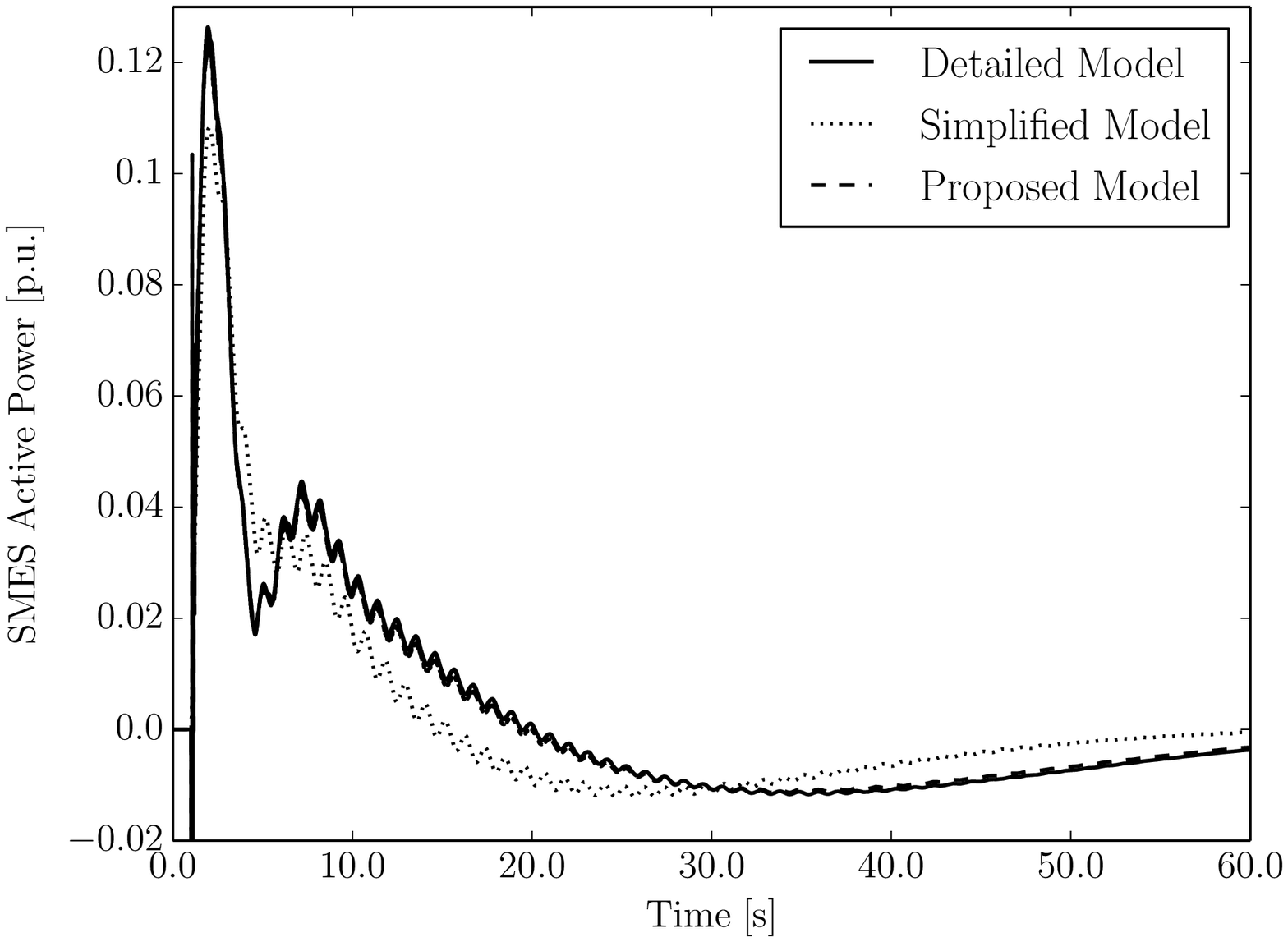}}}
      \label{SMES_Out_noSaturation}  }}
  \subfigure{\parbox[t]{.032\linewidth}{\small (c)\hfill}
    {\parbox[c]{1.03\linewidth}{\resizebox{\linewidth}{!}{\includegraphics{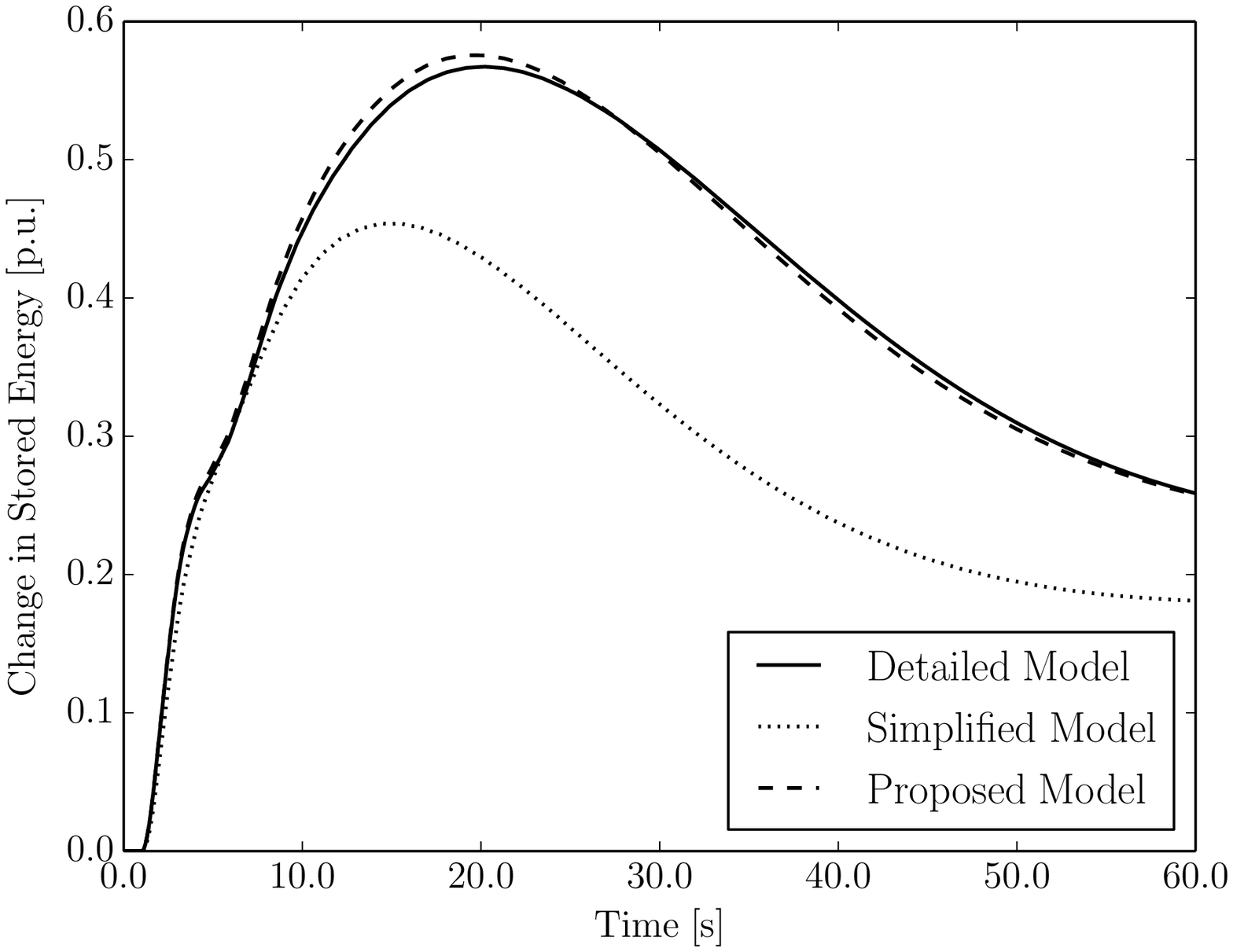}}}
      \label{SMES_Energy_noSaturation} }}
  \caption{Response of the WSCC system with a SMES following a three-phase
    fault at bus 7.  (a) Frequency of the COI.  (b) Active power
    of the SMES. (c) Change in the Stored Energy of the SMES.}
  	\label{SMES_noSat}
\vspace{-0.4cm}
\end{figure}

\begin{figure}[t!]
  \centering
  \subfigure{\parbox[t]{.03\linewidth}{\small (a)\hfill}
    {\parbox[c]{\linewidth}{\resizebox{1.03\linewidth}{!}{\includegraphics{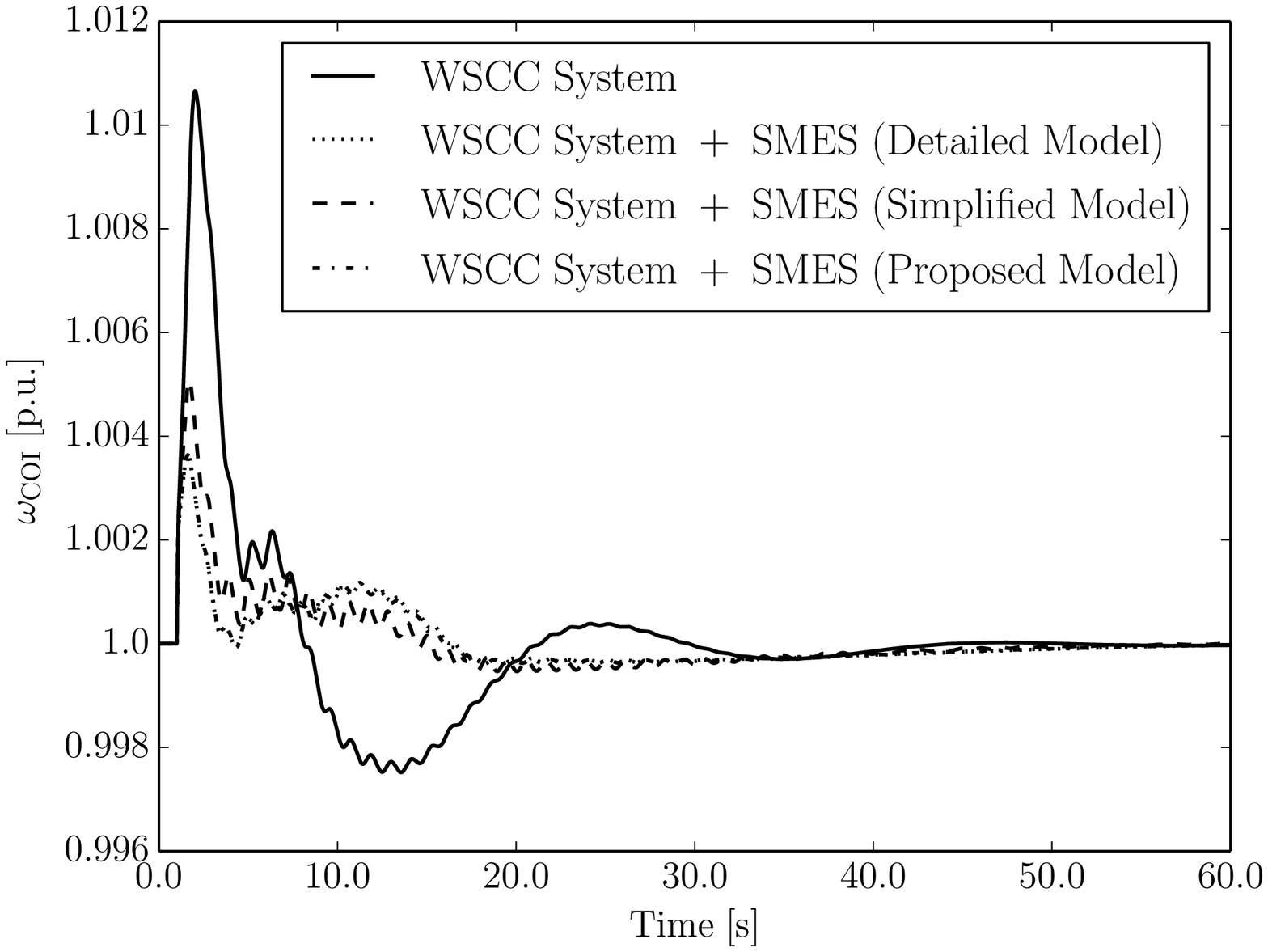}} \vspace{-8mm}}
      \label{SMES_COI_Saturation} }}
  \vspace{-2.5mm}\subfigure{\parbox[t]{.032\linewidth}{\small (b)\hfill}
    {\parbox[c]{\linewidth}{\resizebox{1.03\linewidth}{!}{\includegraphics{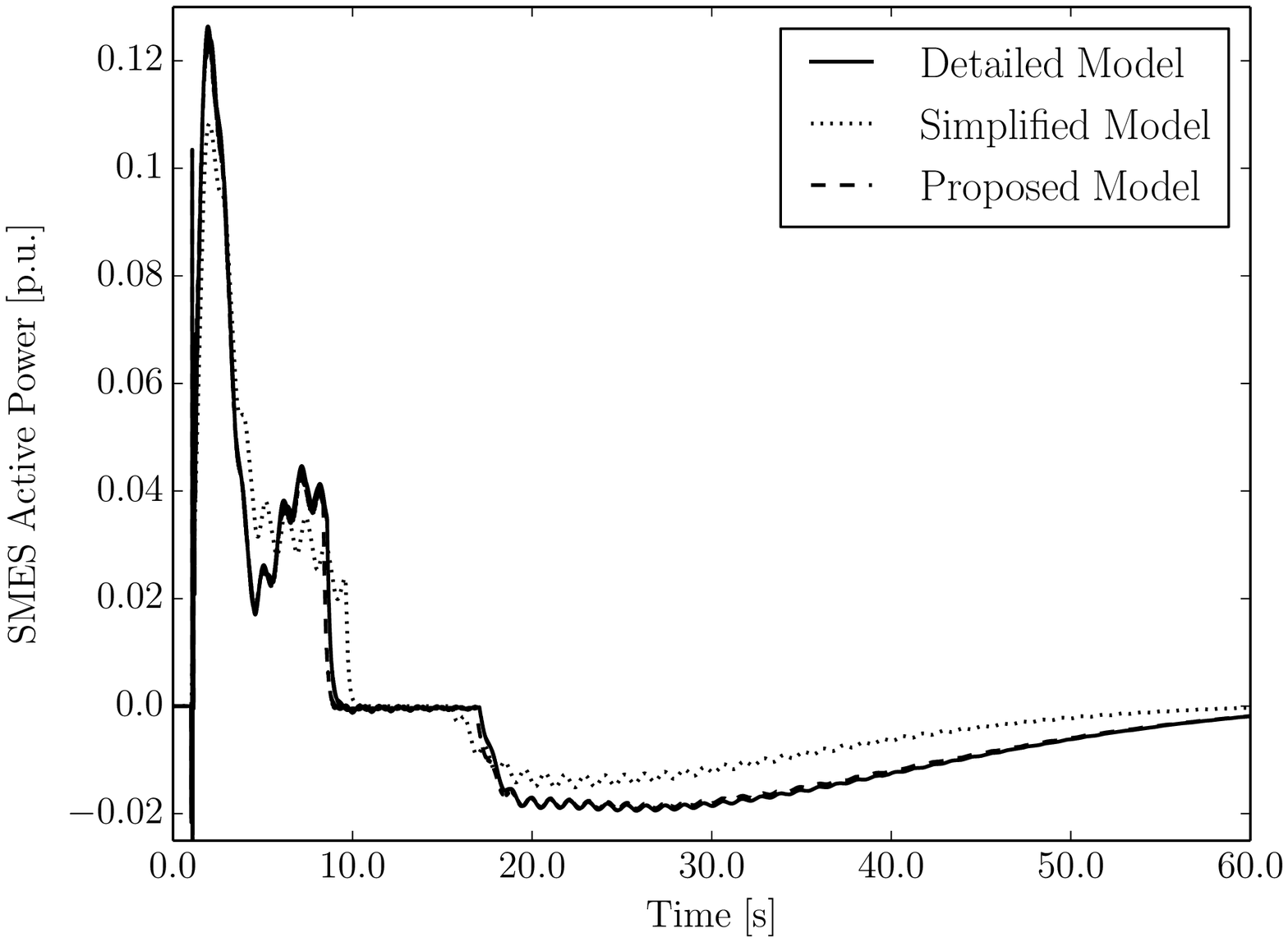}}}
      \label{SMES_Out_Saturation} }}
  \subfigure{\parbox[t]{.032\linewidth}{\small (c)\hfill}
    {\parbox[c]{\linewidth}{\resizebox{1.03\linewidth}{!}{\includegraphics{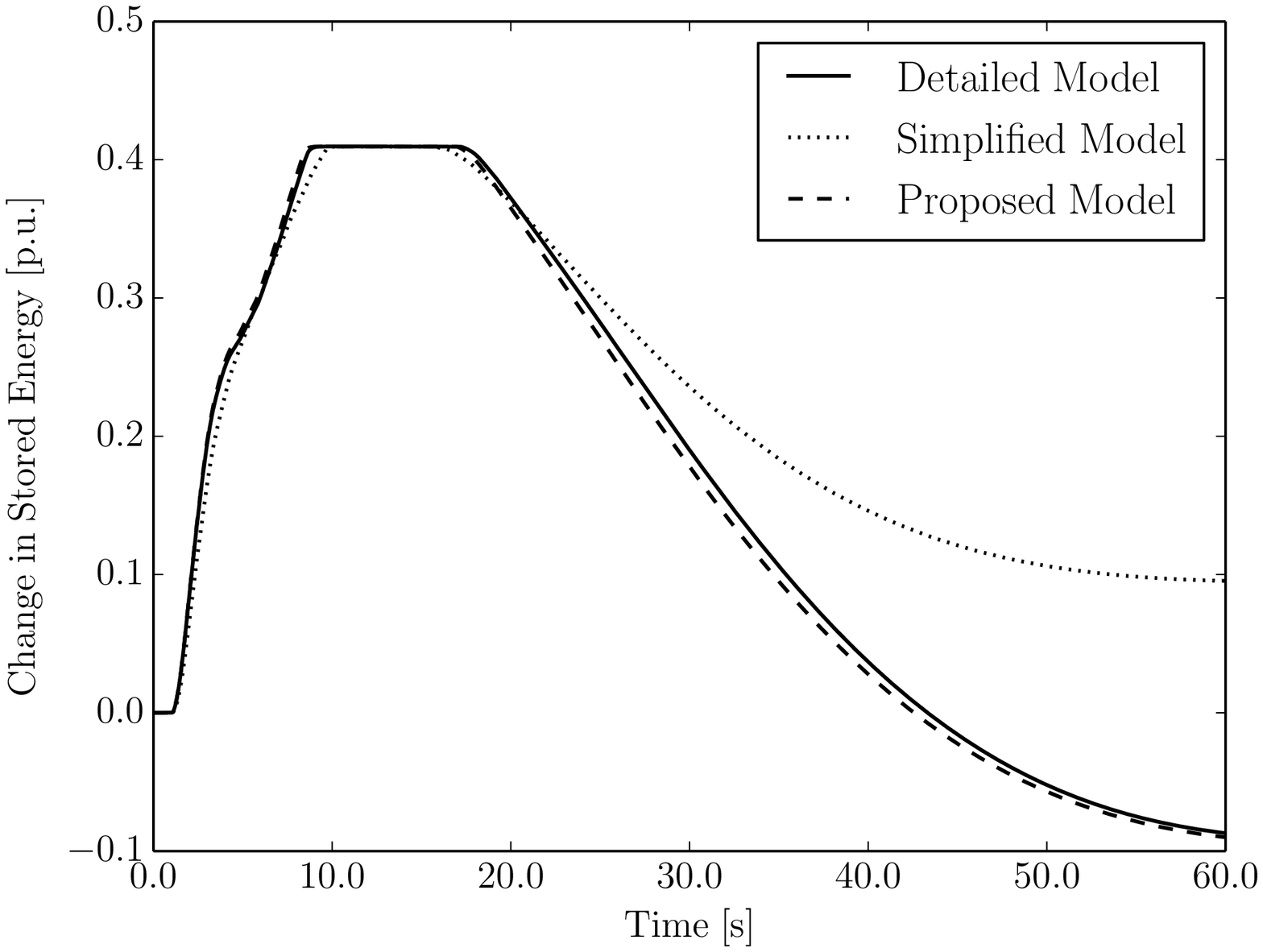}}}
      \label{SMES_Energy_Saturation} }}
  \caption{Response of the WSCC system following a three-phase fault at
    bus 7 when the SMES reaches its maximum storable energy. (a)
    Frequency of the COI. (b) Active power of the SMES. 
    (c) Change in the Stored Energy of the SMES.}
  \label{SMES_Sat}
\vspace{-0.4cm}
\end{figure}
\newpage
\begin{figure*}[t!]
  \centering
  \subfigure{\label{20smes}\parbox[t]{.03\linewidth}{\small (a)\hfill}
    {\parbox[c]{\linewidth}{\resizebox{1.0\linewidth}{!}
    {\includegraphics{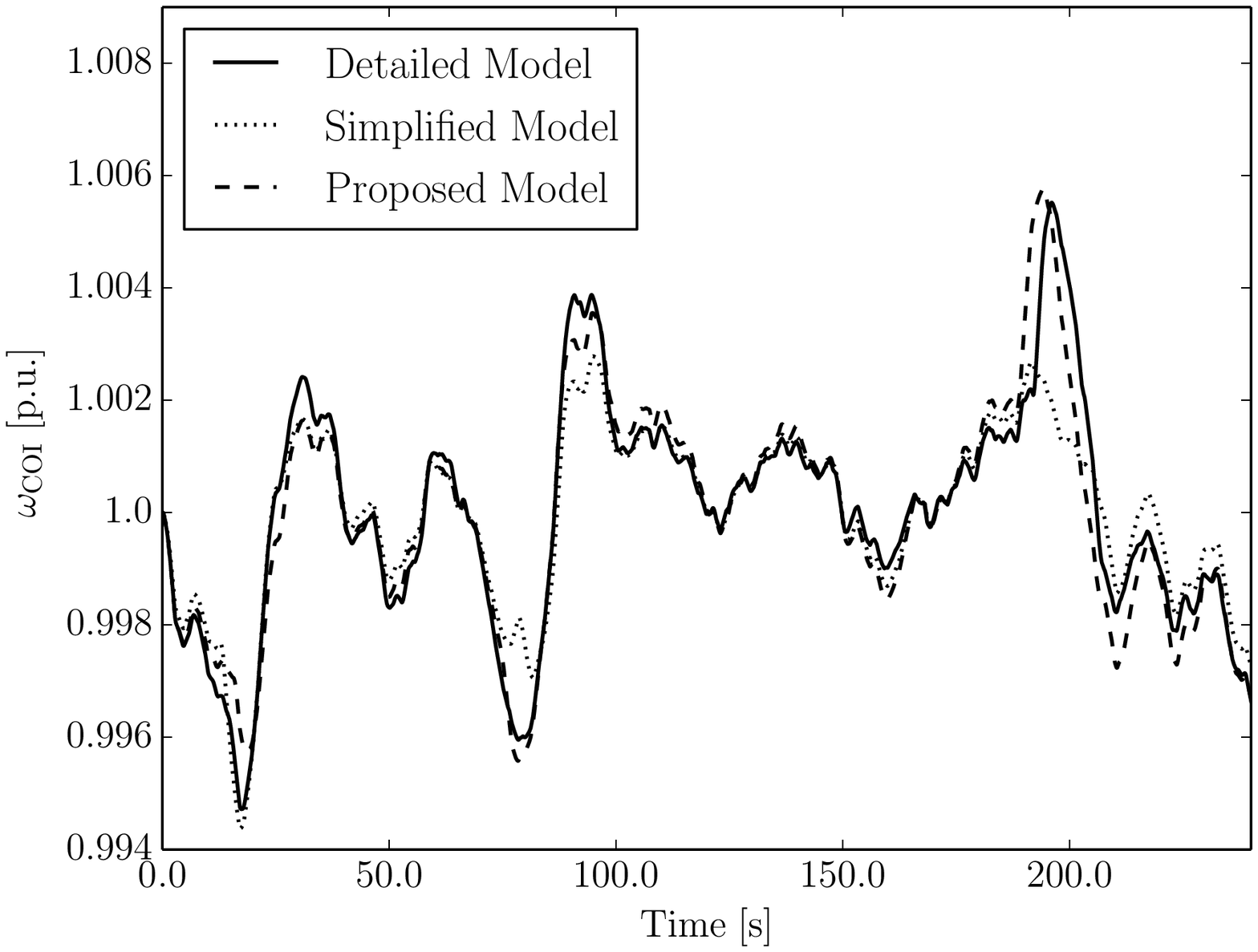}\includegraphics{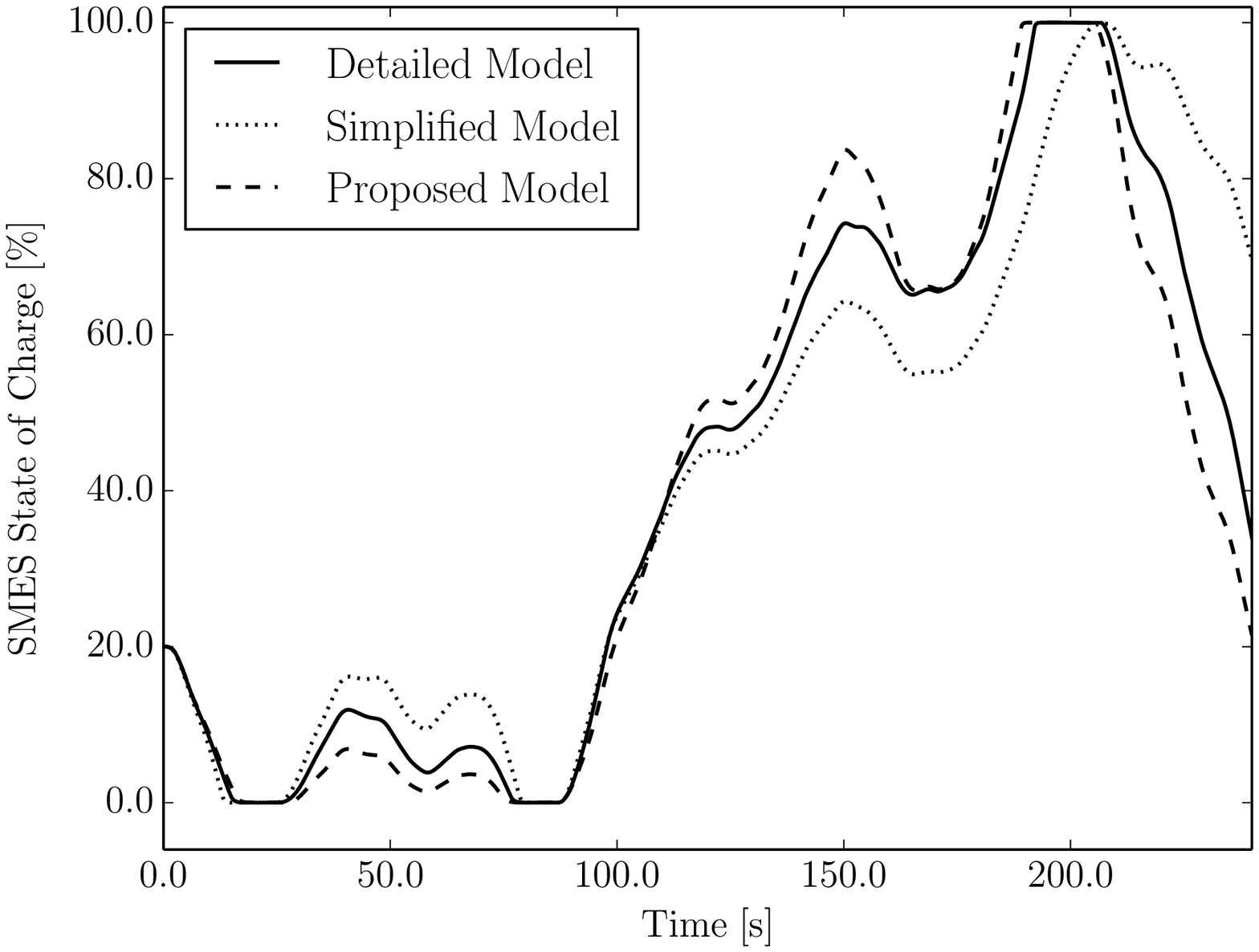}}\vspace*{-0.25cm}}}} 
  \subfigure{\label{50smes}\parbox[t]{.03\linewidth}{\small (b)\hfill}
    {\parbox[c]{\linewidth}{\resizebox{1.0\linewidth}{!}
    {\includegraphics{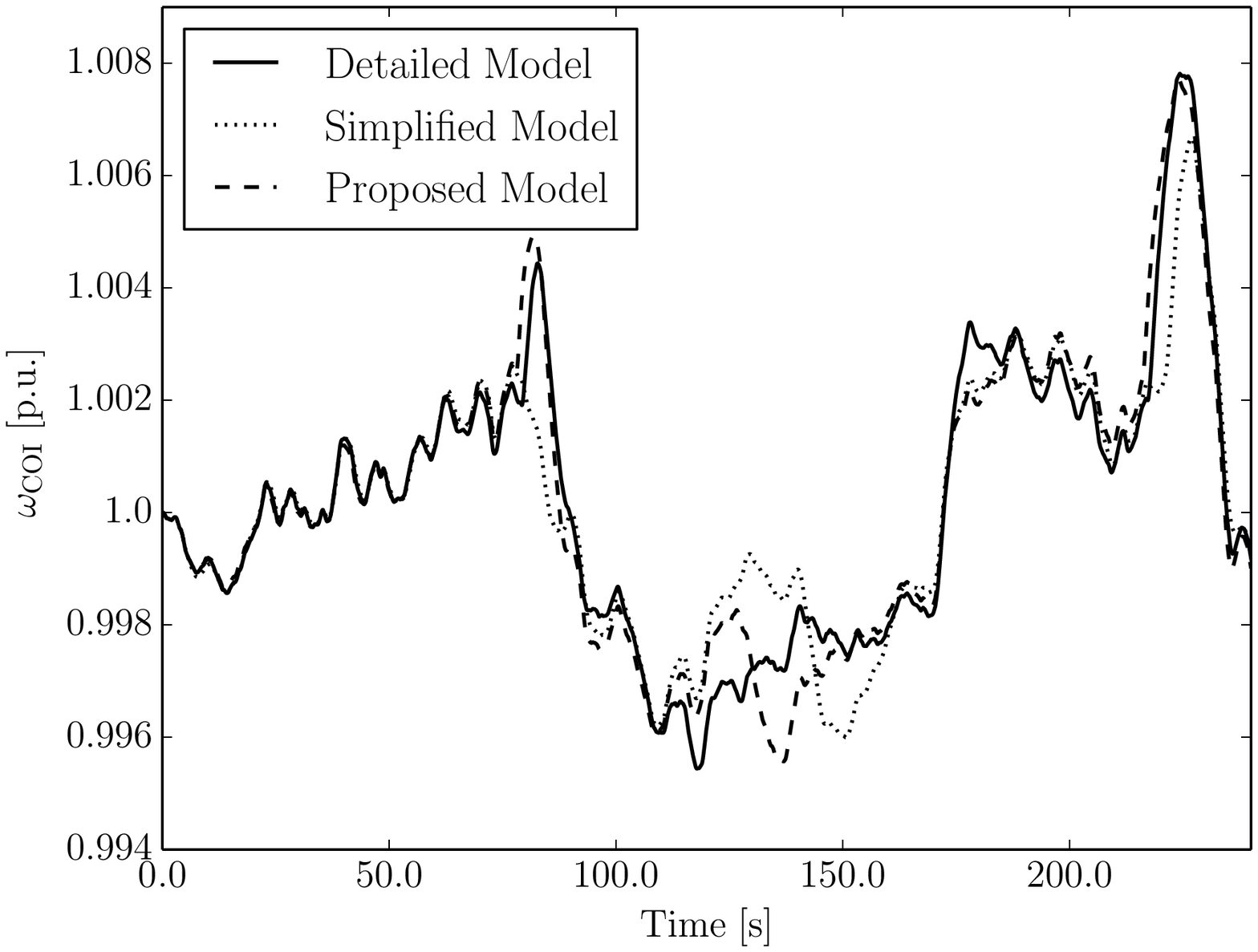}\includegraphics{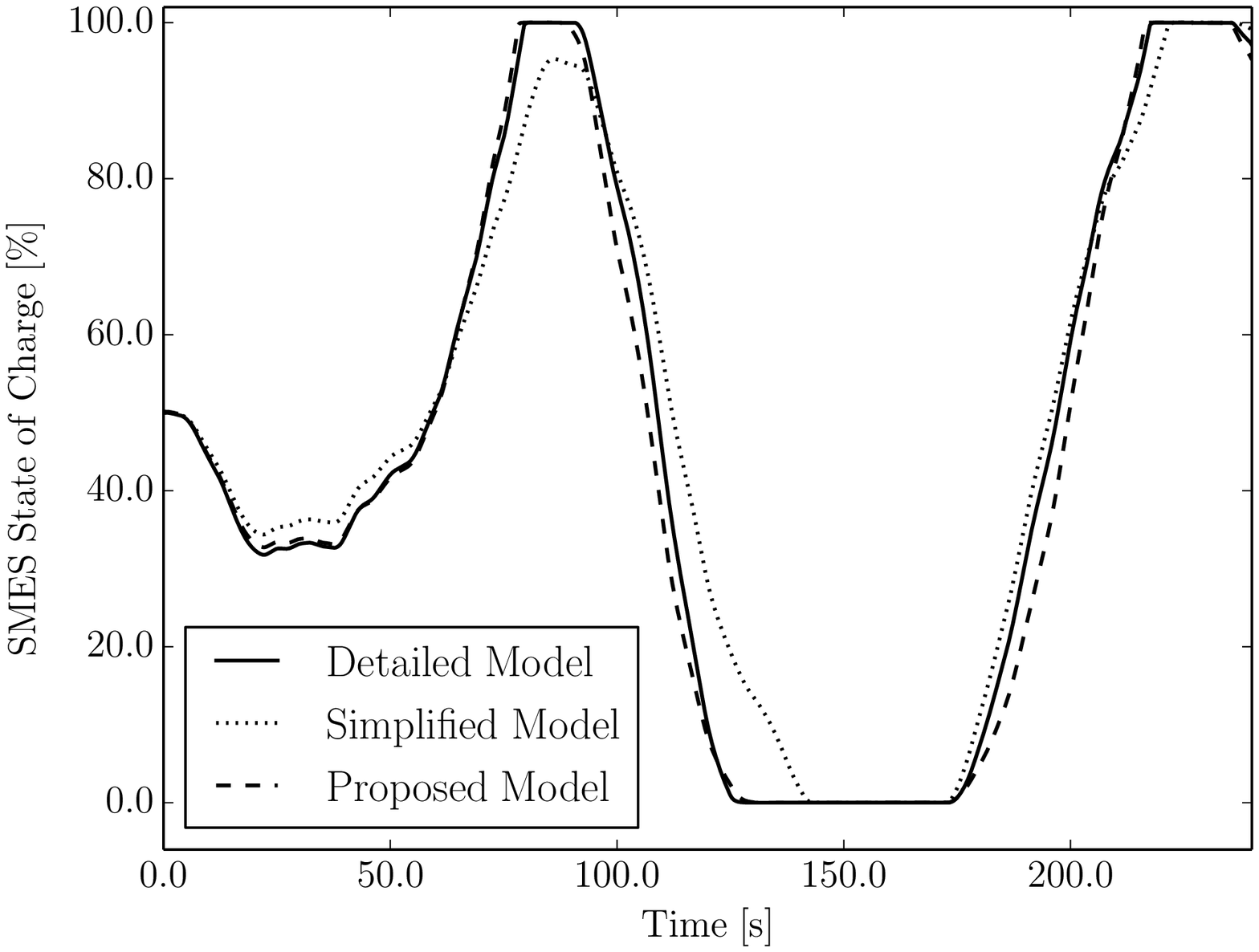}}\vspace*{-0.25cm}}}} 
  \subfigure{\label{80smes}\parbox[t]{.03\linewidth}{\small (c)\hfill}
    {\parbox[c]{\linewidth}{\resizebox{1.0\linewidth}{!}
    {\includegraphics{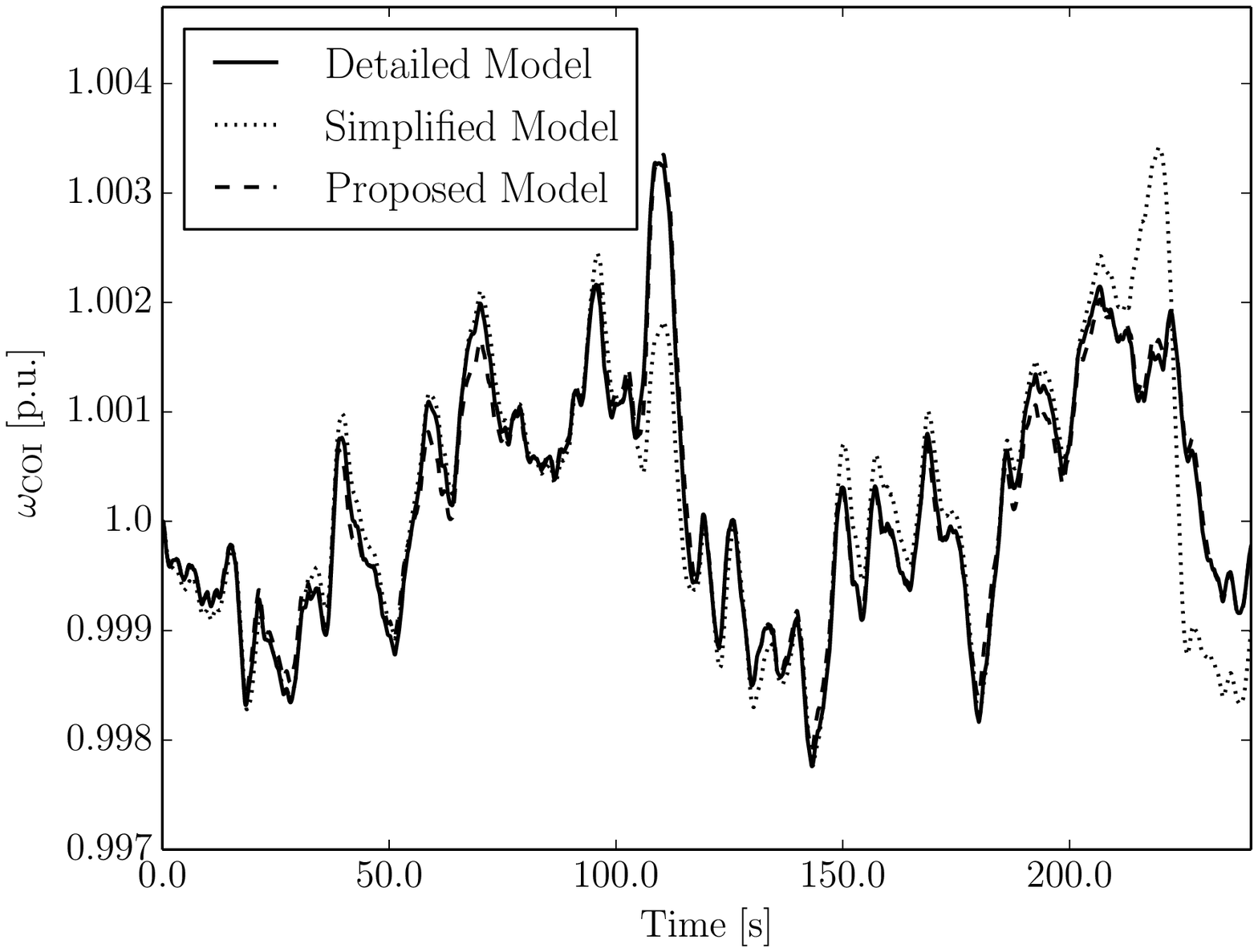}\includegraphics{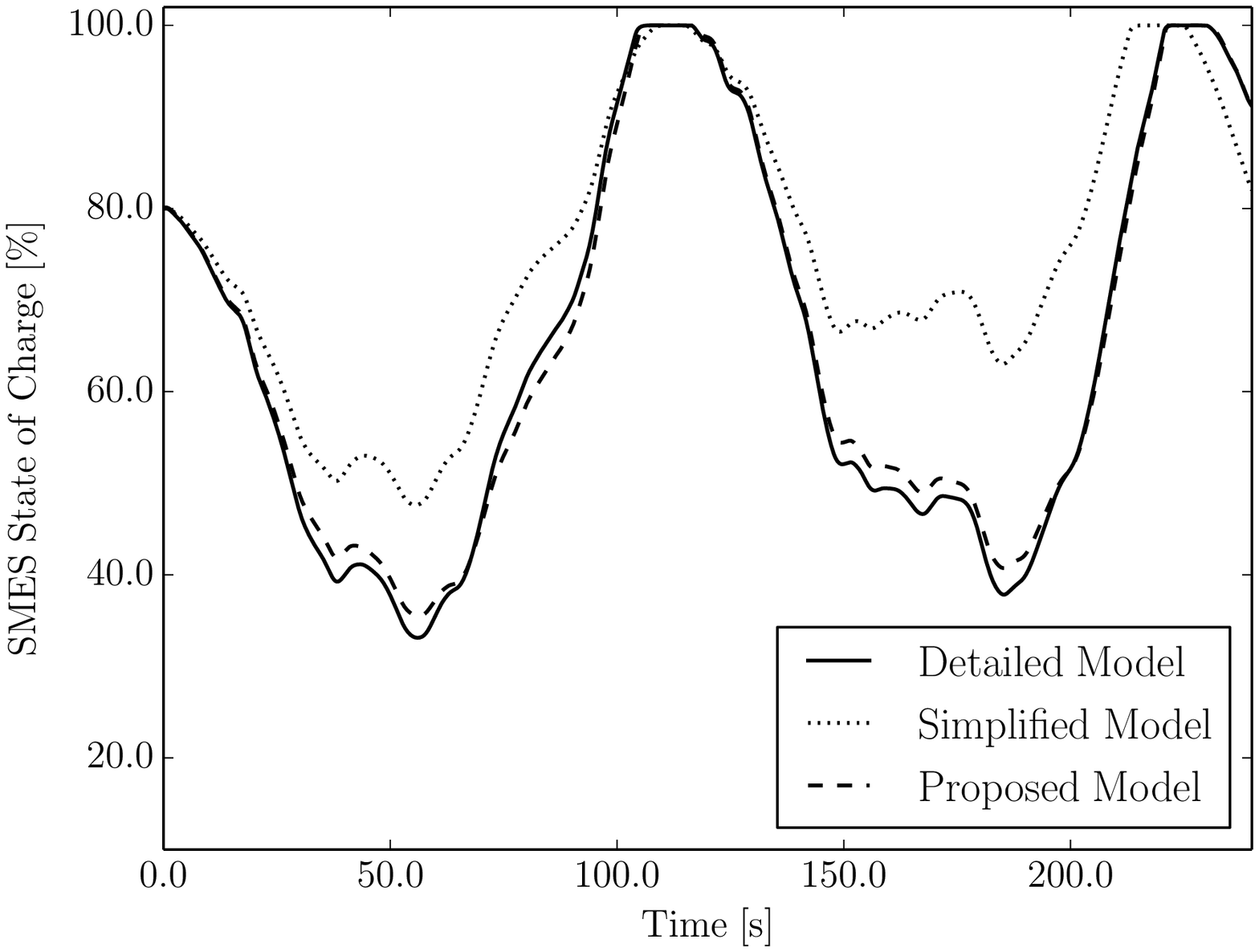}}}}} 
  \caption{Response  of  the  WSCC  system  with  a  SMES  considering
    stochastic variations of the loads.
    (a) The initial state of charge of the SMES is 20\%.
    (b) The initial state of charge of the SMES is 50\%.
    (c) The initial state of charge of the SMES is 80\%.}
  \label{SMES_Stoc}
\vspace{-0.4cm}
\end{figure*}

\subsection{SMES}
\label{subsec:smesTDS}

In this subsection, the storage device connected to bus 8 is a $15$
MW, $60$ MJ SMES, whose model is described in Subsection
\ref{subsec:smes}.  To the best of our knowledge, the largest
  installed SMES (Center of Advanced Power Systems, Florida State
  University) has a capacity of $100$ MJ, and can provide $100$ MW
  peak and $\pm50$ MW oscillatory power~\cite{luongo:03}. 
Two scenarios have been considered: Subsection~\ref{subsub:smesdeterm}
compares the dynamic response of the models of the SMES when the system faces a 
three-phase fault, whereas Subsection~\ref{subsub:smesstoc} considers
stochastic variations of the loads and different initial states of charge of the SMES.

\subsubsection{Contingency}
\label{subsub:smesdeterm}
A three-phase fault occurs at bus 7 at $t = 1$ s and is cleared after
$70$ ms through the disconnection of the line which connects buses 7
and 5.  In this case, the frequency of the center of inertia (COI) of
the system is regulated and its trajectory is shown in
Fig.~\ref{SMES_COI_noSaturation} using the detailed and the proposed
models of the SMES.  Figure \ref{SMES_COI_noSaturation} also shows the
frequency of the COI when the SMES is modeled using the simplified
model described in Subsection \ref{subsec:simpleESS}.

It can be observed that without ESS, the frequency variation after the
fault is around $1\%$, and the steady-state is reached after about
$40$ s.  The inclusion of the SMES in the system allows reducing the
overshoot by $60\%$.  Moreover, the settling time is about $15$ s.

Figure \ref{SMES_Out_noSaturation} compares the active power output of
the SMES when the detailed and the proposed models are used.  The SMES
uses the load notation, therefore positive values of the power
indicate that the SMES is storing energy, and vice versa.

Finally, Fig.~\ref{SMES_Energy_noSaturation} depicts the
  variation of the energy stored in the SMES. The base of the energy
  is 100 MJ, therefore the SMES increases its stored energy a maximum
  of about $57$ MJ during and after the fault.  The steady state value
  of the energy after the occurrence of the disturbance is about $20$
  MJ over the initial conditions.
As it can be observed from Fig.~\ref{SMES_noSat}, the performance of
the system is basically the same when using the detailed and the
proposed models of the SMES.

Figure \ref{SMES_Sat} shows the performance of the system when the
SMES reaches its maximum storable energy. In this simulation, it
  has been considered a maximum variation in the energy of $40$ MJ
  over the initial condition. It can be observed that the effect of
this sort of non-linearity is precisely captured by the proposed
model, and the differences in the performance between the detailed and
the proposed models are very small.

Figures \ref{SMES_noSat} and \ref{SMES_Sat} also show the transient
response of the SMES when the commonly-used simplified model of ESSs described in
Subsection \ref{subsec:simpleESS} is applied.  Gains and time
  constants of the different controllers defined in
  Section~\ref{sec:overview} have the same value for all ESS models.
The trajectories obtained by imposing the simplified model
consistently deviate from the behavior of the detailed one,
particularly after the transient and after reaching the maximum variation of stored
  energy (note that, for a fair comparison, energy limits are also
included in the simplified model in Fig.~\ref{SMES_Sat}). 
The proposed generalized model is able to accurately track the behavior
  of the detailed one in both, fast (transient) and slow
  (post-contingency) dynamics. On the other hand, the simplified model
  is accurate only in the first few seconds after the disturbance.
  The poor performance of the simplified model is an expected
  consequence of the reduced order of its dynamic equations.

\subsubsection{Stochastic Load Variations}
\label{subsub:smesstoc}

In this scenario, all loads are modeled considering stochastic variations
of the power consumptions.
The Ornstein-Uhlenbeck’s process, also known as \it mean-reverting \rm process, is
used to model such variations~\cite{perninge:10, milano:2013}. 
The step size of the Wiener's process is $h = 0.01$s, and the time
step of the time integration $\Delta t$ is set to $0.1$s. Initial values of the load and 
generation are set using a uniform distribution with 5\% of variation with respect to the base-case.

Figure~\ref{SMES_Stoc} shows the frequency of the COI and the state of charge of the SMES 
for three different load profiles and for an initial state of charge of 
20\%, 50\%, and 80\% in Figs.~\ref{20smes},~\ref{50smes}, and~\ref{80smes}, respectively.
These percentages are expressed in terms of the allowed energy 
variability of the storage device.
For each simulation, SMES models are simulated considering same load variation profiles. 

The following remarks are relevant:
\begin{itemize}
\item[i. ] The proposed generalized ESS model tracks the beha\-vior of the detailed one better than
the commonly-used simplified model of the SMES, for any initial state of charge.
\item[ii. ] The proposed model is more accurate the closer is the state of charge to the initial condition
during the simulations. This is an expected property of linearized models.
\item[iii. ] Based on the results of hundreds of simulations with different load profiles and initial 
conditions, it has been observed that the best average accuracy of the proposed
generalized model is obtained by computing the matrices of (\ref{GenModel})-(\ref{GenEnergy}) for an
initial state of charge of 50\%.
\end{itemize}


\subsection{CAES}
\label{subsec:caesTDS}

In this case study, we consider a $15$ MW, $30$ bar CAES, whose
model is described in Subsection \ref{subsec:caes}. The CAES regulates
the active power flowing through the transformer connecting buses 2
and 7.  As in previous Subsection, the CAES uses the load
notation, and two scenarios are performed:
Subsection~\ref{subsub:caesdeterm} considers a loss of load as contingency,
while Subsection~\ref{subsub:caesstoc} includes stochastic variations in all loads.
Note that the maximum size of currently installed
above-ground CAESs is up to $5$ MW \cite{ibrahim:08}.  In this
example, we assume that the CAES has three times this capacity. \\

\subsubsection{Contingency}
\label{subsub:caesdeterm}
\begin{figure}[t!]
  \centering
  \subfigure{\parbox[t]{.03\linewidth}{\small (a)\hfill}
    {\parbox[c]{\linewidth}{\resizebox{1.0\linewidth}{!}{\includegraphics{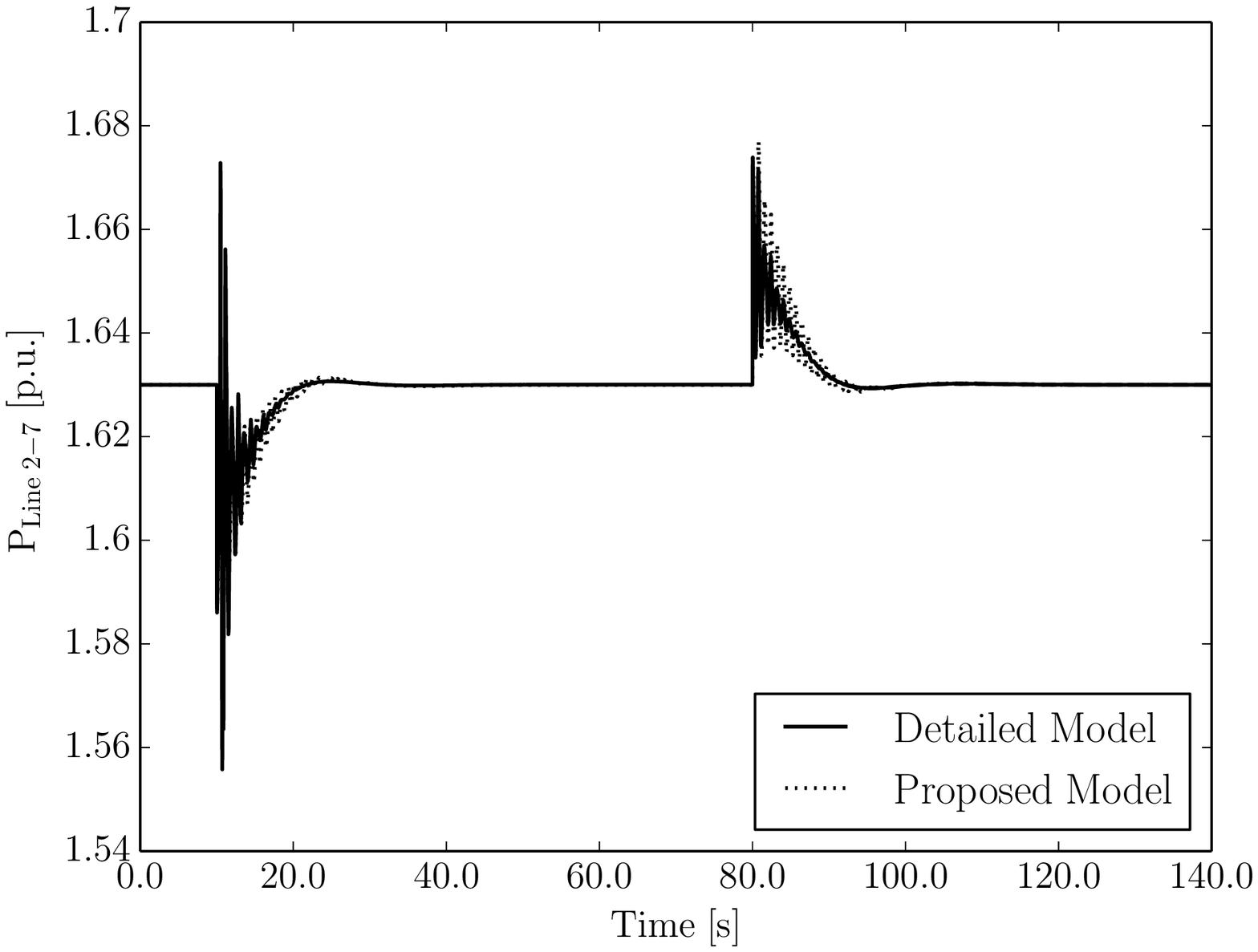}} \vspace{-3mm}}
    \label{CAES_line7} }}\vspace{-0.4cm}
  \subfigure{\parbox[t]{.032\linewidth}{\small (b)\hfill}
    {\parbox[c]{\linewidth}{\resizebox{1.0\linewidth}{!}{\includegraphics{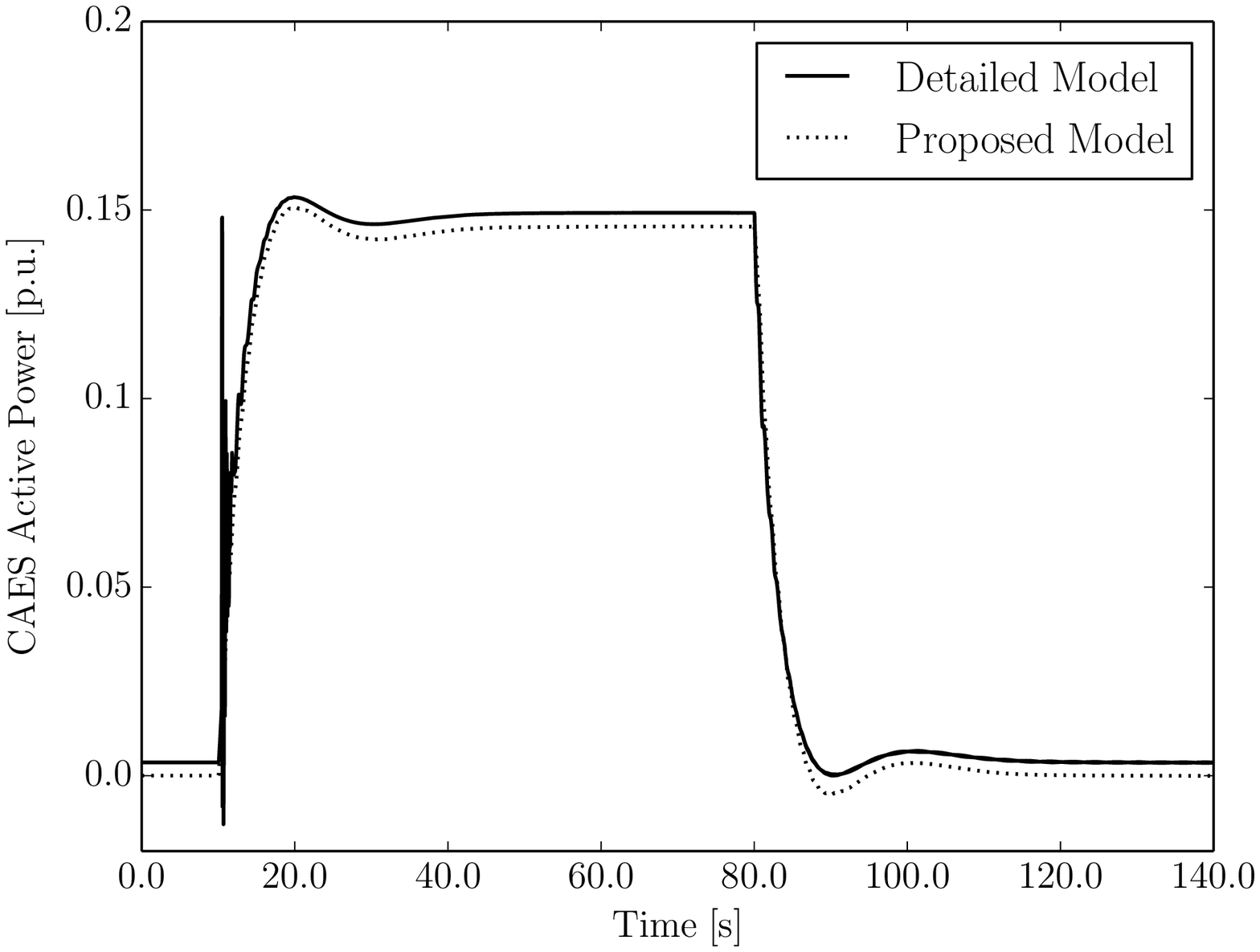}}}
    \label{CAES_COI} }}
  \caption{Response of the WSCC system with a CAES following a loss of
    load. (a) Power flowing through the transformer connecting buses 2 and 7. (b)
    Active power of the CAES.}
\vspace{-0.4cm}
\end{figure}

The contingency is a loss of load of $15$ MW occurring
at bus 5 at $t = 10$ s.  Finally, the load is reconnected at $t = 80$
s.  Figures \ref{CAES_line7} and \ref{CAES_COI} illustrate the active
power flowing through the transformer connecting buses 2 and 7 and the
active power of the CAES, respectively. It can be seen
  that the response of the CAES simulated using the proposed model is
  accurate despite the complexity of the detailed CAES model shown in
  Subsection~\ref{subsec:caes}.

\begin{figure}[h!]
  \centering
  \subfigure{\parbox[t]{.03\linewidth}{\small (a)\hfill}
    {\parbox[c]{\linewidth}{\resizebox{1.0\linewidth}{!}{\includegraphics{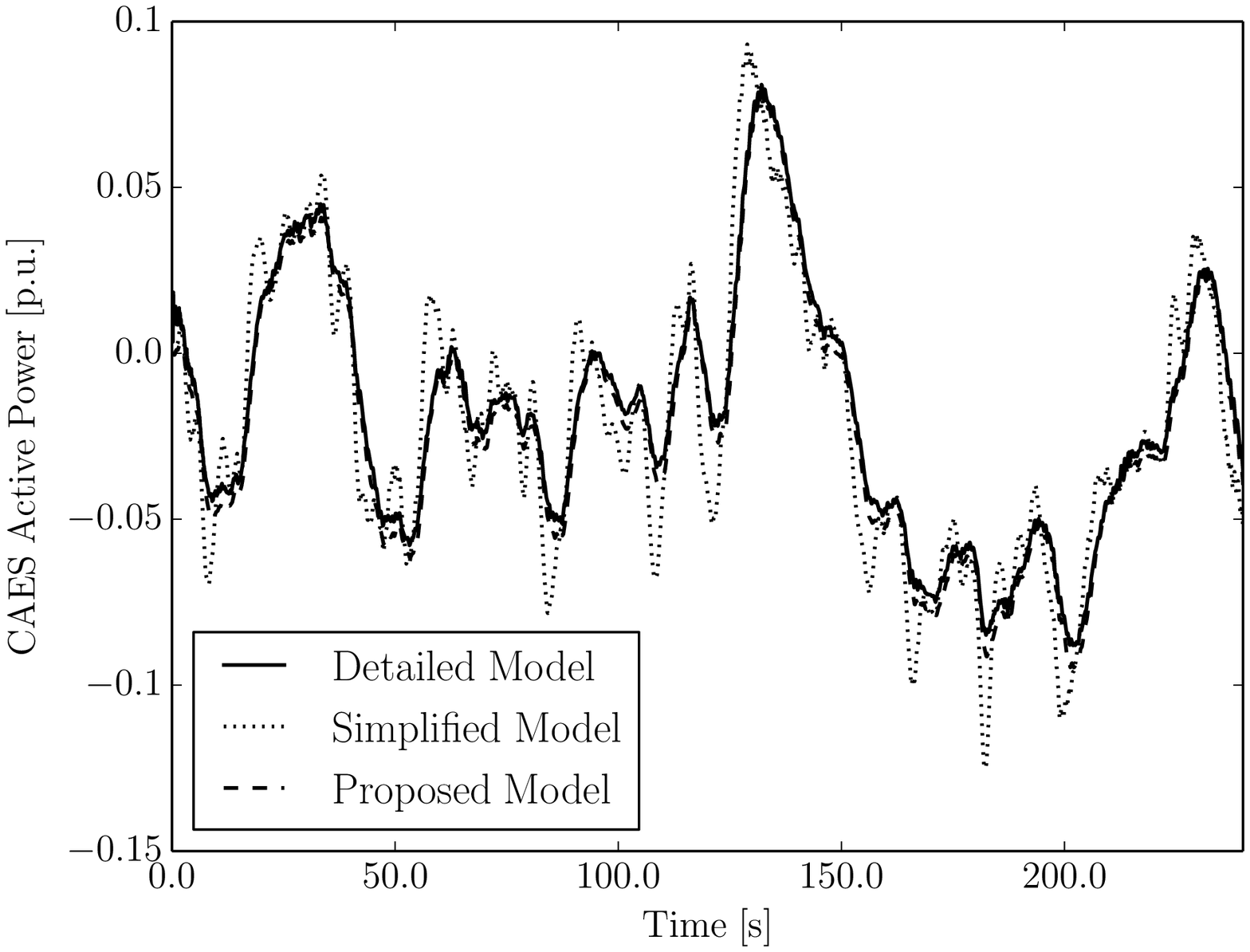}} \vspace{-3mm}}
    \label{CAES_Stoc} }}\vspace{-0.4cm}
  \subfigure{\parbox[t]{.032\linewidth}{\small (b)\hfill}
    {\parbox[c]{\linewidth}{\resizebox{1.0\linewidth}{!}{\includegraphics{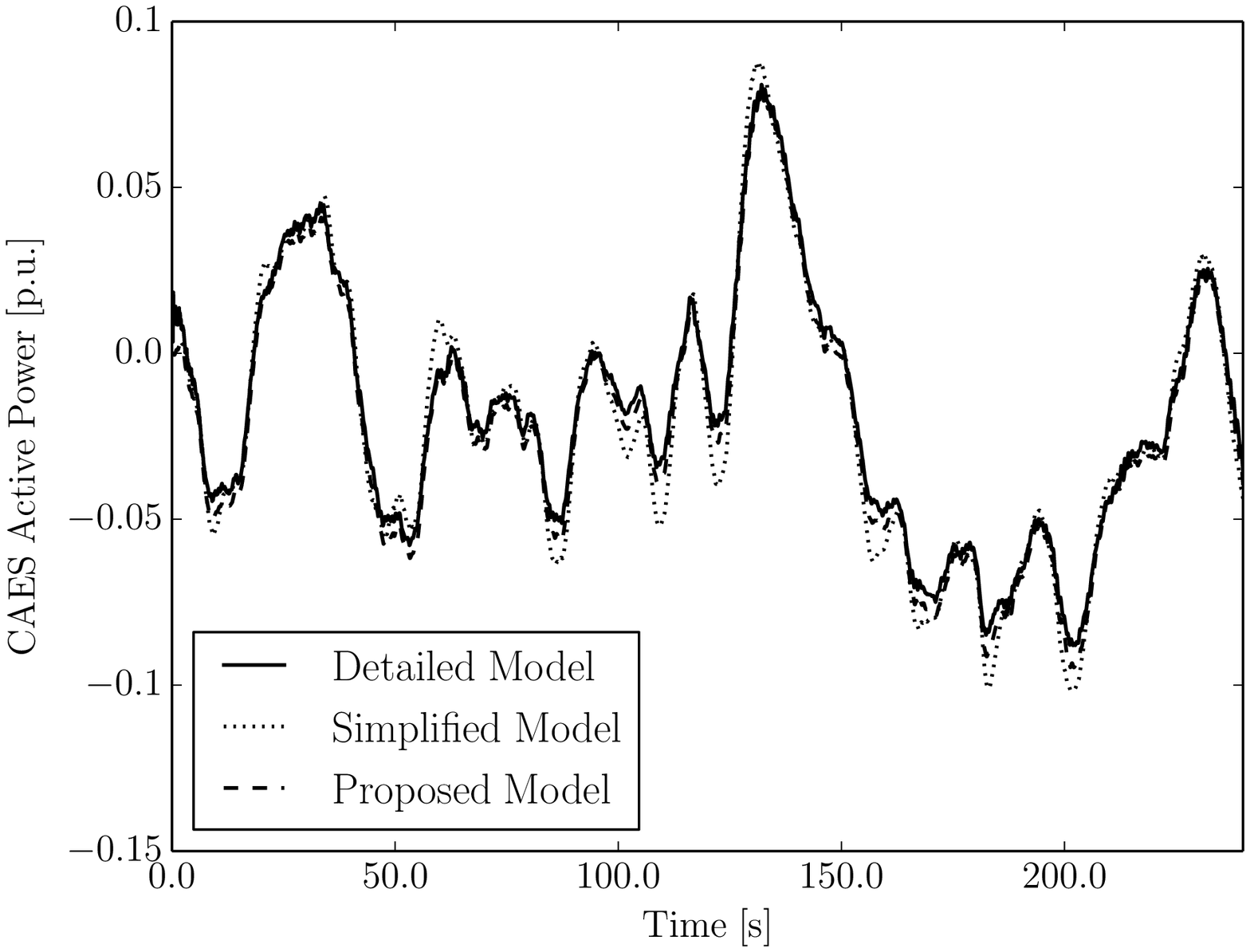}}}
    \label{CAES_Stoc_tuned} }}
  \caption{Response of the WSCC system with a CAES considering
    stochastic perturbations in the loads.
    (a) Same control parameters for all models.
    (b) Control parameters of the simplified CAES model are three times 
		    smaller than those of the detailed and the generalized ones.}
\vspace{-0.4cm}
\end{figure}

\subsubsection{Stochastic Load Variations}
\label{subsub:caesstoc}
In this scenario, all loads are modeled using the \it mean-reverting \rm process,
(see Subsection~\ref{subsub:smesstoc}). Figure~\ref{CAES_Stoc}
depicts the active power of the CAES for each model of this device, for an initial state 
of charge of about 50\%. All control parameters are the same for all models. 
The time constants of the simplified model of the CAES have been tuned after several
trial and error attempts, in order to obtain the behavior as close as possible to 
the detailed model. Figure~\ref{CAES_Stoc} shows that the response of the
commonly-used simplified model differs considerably from the one of the detailed model.
On the other hand, the proposed generalized model appears to be very accurate.
In order to obtain a more accurate response of the storage device using
the simplified model, it is required to properly tune the control gains $K_{\rm p \it P}$ 
and $K_{\rm i \it P}$ of the controller depicted in Fig.~\ref{simplescheme}. 
Figure~\ref{CAES_Stoc_tuned} depicts the response of the CAES considering the same 
load profiles as in Fig.~\ref{CAES_Stoc}, and when the gains $K_{\rm p \it P}$ 
and $K_{\rm i \it P}$ are three times smaller than the gains $K_{\rm pu}$ and
$K_{\rm iu}$ of the storage controller of Fig.~\ref{inputsignalcontrol}, respectively.
Some remarks can be deduced:
\begin{itemize}
\item[i.]   The accuracy  of the  proposed generalized  model 
  does not appear to be affected by the complexity and the size of order 
  reduction of the original detailed model. Note that the order reduction
  achieved in the case of the CAES is from 29 variables in the detailed model
  to only 5 in the proposed one. 
\item[ii.] The  accuracy of the commonly-used  simplified model cannot
  be guaranteed even  if a lengthy and careful tuning  of its control
  parameters is carried out.
\item[iii.]  Because  of the  required tuning,  the design  of control
  strategies cannot  be based on  the simplified model.  On  the other
  hand,  exactly same  control parameters  can  be used  for both  the
  generalized and detailed  ESS models.  
\end{itemize}


\vspace{-2mm}

\subsection{BES}
\label{subsec:bessTDS}

In this example, a $40$ MW BES is installed at bus 8~\cite{shepherd:65}.
The data of the BES is taken from~\cite{mercier:09}, 
that describes a $55$ MW, $76.7$ GJ BES used for frequency control.
The contingency is a loss of a $40$
MW load at bus 5.  The load is disconnected at $t = 10$ s, and is
reconnected at $t = 100$ s.  In this case study, the BES regulates the
frequency of the COI.

\begin{figure}[t!]
  \centering
  \subfigure{\parbox[t]{.03\linewidth}{\small (a)\hfill}
    {\parbox[c]{\linewidth}{\resizebox{1.0\linewidth}{!}{\includegraphics{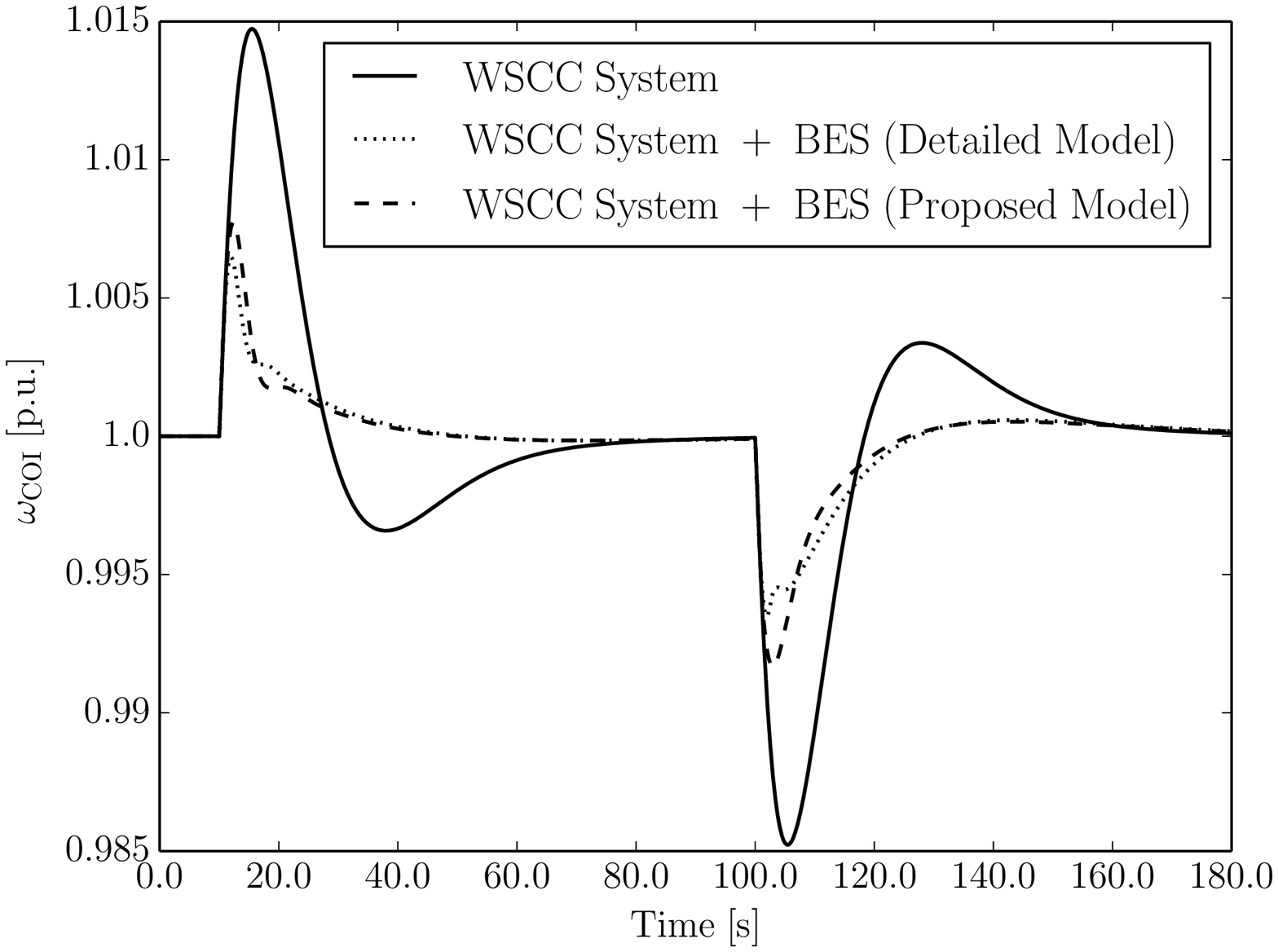}} \vspace{-3mm}}
    \label{BESS_COI}  }}\vspace{-0.4cm}
  \subfigure{\parbox[t]{.03\linewidth}{\small (b)\hfill}
    {\parbox[c]{\linewidth}{\resizebox{1.0\linewidth}{!}{\includegraphics{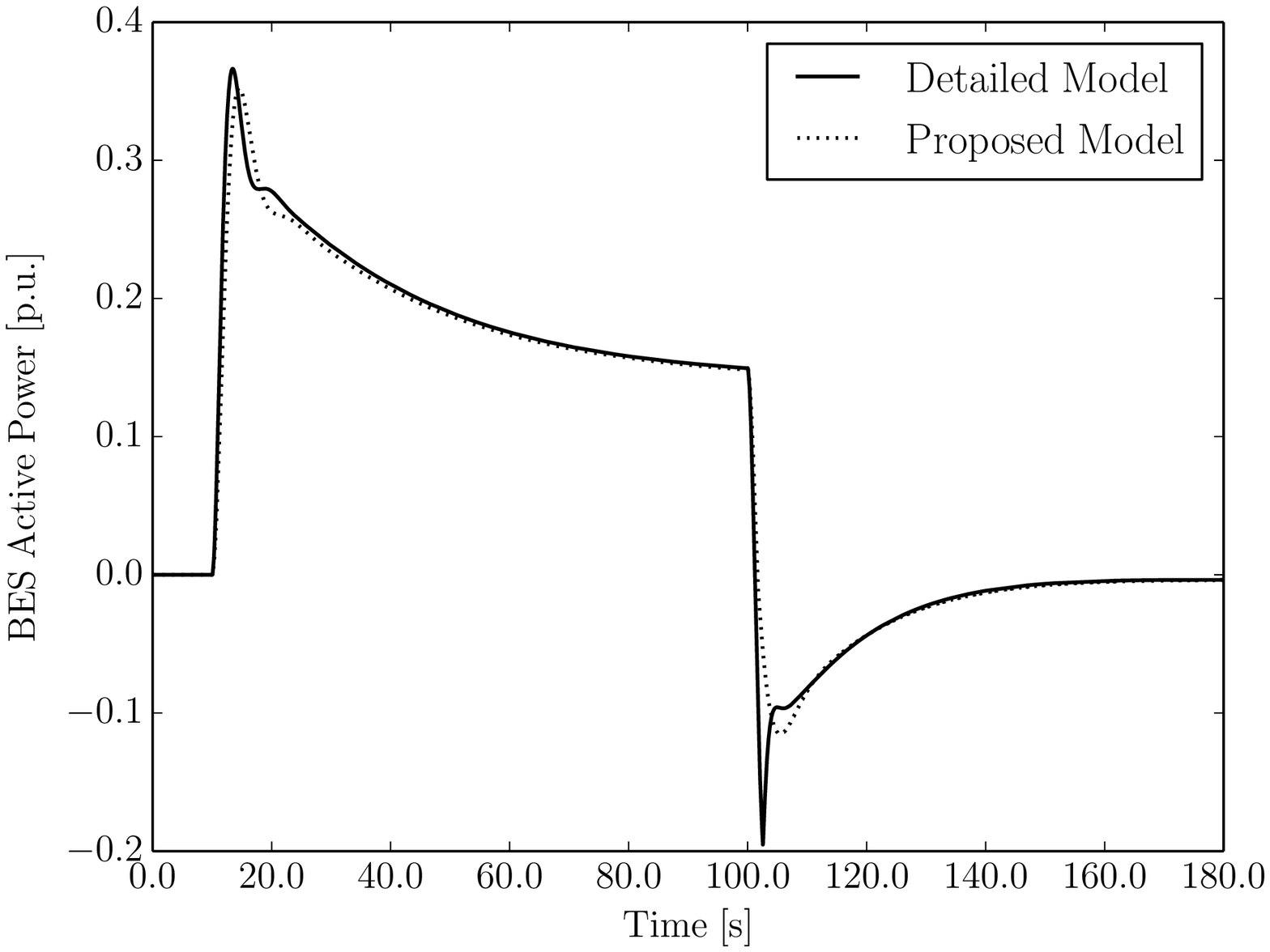}}} }
    \label{BESS_Out} }
  \caption{Response of the WSCC system with a BES following a loss of
    load. (a) Frequency of the COI. (b) Active power of the
    BES.} \label{BESScasestudy}
\end{figure}

Figures \ref{BESS_COI} and \ref{BESS_Out} show the frequency of the
COI and the power output of the BES, respectively. The initial
  SOC of the battery is set to $85\%$ to force operating close to the
  nonlinear voltage-current characteristic of the battery
  \cite{tremblay:07, shepherd:65}.  
As stated in Subsection
\ref{subsec:bess}, two sets of matrices for the general model 
of the BES have to be considered
because of the discontinuity in the polarization voltage, $v_{\rm p}$,
depending on whether the battery is charging (time from $t = 10$ s to $t = 100$ s)
or discharging (time from $t = 100$ s). It can be seen from Fig.~\ref{BESScasestudy}
that the generalized model is able to track, with a good level of accuracy, 
the behavior of the detailed model of the BES for both operating conditions, despite
the nonlinearities of the device at such a high initial SOC, and the switching
between the two operating modes.

\subsection{Concluding Remarks}
\label{sec:remarks}

The following remarks on the proposed generalized model of ESSs are
relevant.

\begin{enumerate}
\item \textit{The proposed model provides a good compromise bet\-ween
    simplicity and accuracy.} In fact, the proposed model has a
  reduced and constant dynamic order but, nevertheless it can
  reproduce faithfully the behavior of ESSs whose detailed
  transient stability models are
  considerably more complex (e.g., CAES).  On the contrary, the case
  study shows that simpler models of ESSs might not be precise,
  particularly if windup limiters are binding.
\item \textit{Linearization of a subset of ESS equations does not
  affect accuracy.}  The rationale behind this observation is that
  most ESS variables involved in nonlinear equations are bounded
  and, hence, the ESS operating point does not vary consistently even
  after a large perturbation.
\item \textit{Detailed ESS models are needed to define the para\-meters
  of the proposed model.}  These data have to be provided by ESS
  manufactures, in the same way makers provide $d$- and $q$-axis
  reactances and time constants of the synchronous machine Park model.
  Hence, we consider that the proposed ESS model can be an opportunity
  to define a ESS standard for transient stability analysis.
\item \textit{Control strategies designed for the proposed generalized model
	can be directly applied to the detailed ESS models.} The linear structure of
	the equations of the proposed model simplifies the design of more advanced 
	and robust control strategies for ESSs that could be applied subsequently
	to the detailed models of these devices.
\end{enumerate}


\section{Conclusions}
\label{sec:conclu}

This paper proposes a generalized model of energy storage devices for
voltage and angle stability analysis.  Such a model allows simulating
different technologies using a fixed set of DAEs and parameters.
Simulation results show that the proposed model is able to accurately
reproduce the dynamic behavior of detailed transient
stability models for large disturbances, namely faults
and loss of loads, and across their whole operating cycle. 
The non-linearity of ESS controllers, i.e., hard
limits, are also properly taken into account by the proposed model.

The proposed generalized ESS model appears to be particularly useful to
synthesize and compare different control strategies.  In fact, since
the proposed model is composed of a fixed set of equations, the same
control scheme can be straightforwardly used for testing the dynamic
response of different technologies. This will be addressed in future
work.

\section*{Acknowledgment}

This work was conducted in the Electricity Research Centre, 
University College Dublin, Ireland, which is supported by the 
Electricity Research Centre’s Industry Affiliates Programme 
(http://erc.ucd.ie/industry/). This material is based upon works 
supported by the Science Foundation Ireland, by funding Alvaro Ortega 
and Federico Milano, under Grant No. SFI/09/SRC/E1780. The opinions, 
findings and conclusions or recommendations expressed in this material 
are those of the authors and do not necessarily reflect the views of the 
Science Foundation Ireland. The second author is co-funded by EC Marie 
Sk{\l}odowska-Curie Career Integration Grant No. PCIG14-GA-2013-630811.

\bibliographystyle{IEEEtran}
\bibliography{references}


\begin{biography}
  [{\includegraphics[width=1in,height=1.25in,clip,keepaspectratio]{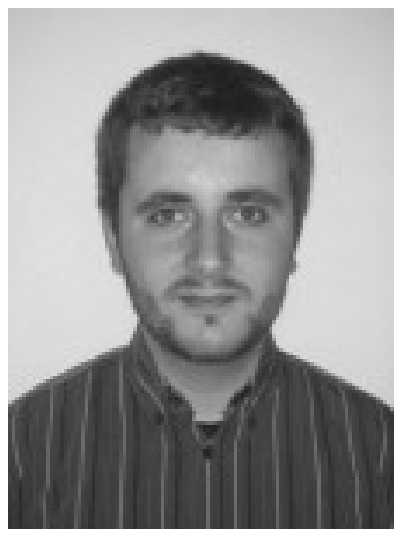}}]
  {{\'Alvaro Ortega}} (S'14) received from Escuela Superior de
  Ingenieros Industriales, University of Castilla-La Mancha, Ciudad
  Real, Spain, the degree in Industrial Engineering in 2013, with a
  final project on modeling and dynamic analysis of compressed air
  energy storage systems.  Since September 2013, he is a Ph.D. student
  candidate with the Electricity Research Centre, University College
  Dublin, Ireland.
\end{biography}


\begin{biography}
  [{\includegraphics[width=1in,height=1.25in,clip,keepaspectratio]{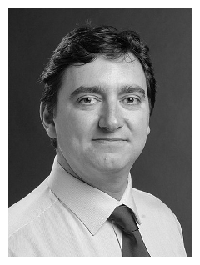}}]
  {Federico Milano} (SM'09) received from the University of Genoa,
  Italy, the Electrical Engineering degree and the Ph.D.~degree in
  1999 and 2003, respectively.  From 2001 to 2002 he was with the
  University of Waterloo, Canada, as a Visiting Scholar.  From 2003 to
  2013, he was with the University of Castilla-La Mancha, Ciudad Real,
  Spain.  In 2013, he joined the University College Dublin, Ireland,
  where he is currently an associate professor.  His research
  interests include power system modeling, stability and control.
\end{biography}

\end{document}